\documentclass[journal]{IEEEtran}
\usepackage{multirow}
\usepackage{color}
\usepackage{indentfirst}
\usepackage{amsmath}
\ifCLASSINFOpdf
  \usepackage[pdftex]{graphicx}
  \usepackage[justification=centering]{caption}
  \graphicspath{{./figures/}{./biographies/}}
  \DeclareGraphicsExtensions{.pdf,.jpeg,.png,.jpg}
\else
\fi

\ifCLASSOPTIONcompsoc
  \usepackage[caption=false,font=normalsize,labelfont=sf,textfont=sf]{subfig}
\else
  \usepackage[caption=false,font=footnotesize]{subfig}
\fi
\usepackage{algorithm}
\usepackage{algorithmic}

\usepackage{cite}

\begin{document}
%
\title{Beyond Camera Motion Blur Removing: How to Handle Outliers in Deblurring }
%
%
%

\author{Meng Chang*,
        Chenwei Yang*,
        Huajun Feng,
        Zhihai Xu,
        and~Qi~Li
\thanks{M Chang, C Yang, H Feng, Z Xu and Q Li are with the State Key Laboratory of Modern Optical Instrumentation, Zhejiang University, Hangzhou, Zhejiang, 310027 China e-mail:liqi@zju.edu.cn.}

\thanks{* The authors contribute equally.}

}

%
%

\markboth{IEEE TRANSACTIONS ON COMPUTATIONAL IMAGING, VOL. 6, 2020}%
{Shell \MakeLowercase{\textit{et al.}}: Bare Demo of IEEEtran.cls for IEEE Journals}
%



\maketitle

\begin{abstract}
Camera motion deblurring is an important low-level vision task for achieving better imaging quality. When a scene has outliers such as saturated pixels, the captured blurred image becomes more difficult to restore. In this paper, we propose a novel method to handle camera motion blur with outliers. We first propose an edge-aware scale-recurrent network (EASRN) to conduct deblurring. EASRN has a separate deblurring module that removes blur at multiple scales and an upsampling module that fuses different input scales. Then a salient edge detection network is proposed to supervise the training process and constraint the edges restoration. By simulating camera motion and adding various light sources, we can generate blurred images with saturation cutoff. Using the proposed data generation method, our network can learn to deal with outliers effectively. We evaluate our method on public test datasets including the GoPro dataset, Kohler’s dataset and Lai’s dataset. Both objective evaluation indexes and subjective visualization show that our method results in better deblurring quality than other state-of-the-art approaches.
\end{abstract}

\begin{IEEEkeywords}
deblurring, outlier removal, image restoration, image processing.
\end{IEEEkeywords}

%
\IEEEpeerreviewmaketitle

\section{Introduction}
%
%
%
%
\IEEEPARstart{I}{mage} blur due to camera motion is a common type of image quality degradation. Imaging system aberrations, defocusing, camera motion during exposure and subject motion in scenes are the four main factors that lead to image blur. Unlike the other factors—which may be the photographer’s intent—camera motion blur is generally not expected. Camera motion deblurring benefits include the ability to extend the shutter duration during photography and image quality improvements, such as higher dynamic range, more accurate color reproduction, and lower noise levels.

Image deblurring is a fundamental research area in image restoration. In recent decades, the traditional deblurring research approaches have focused on three aspects, i.e., the noise probability model~\cite{shan2008high, cho2011handling, whyte2014deblurring, bar2006image}, image context priors~\cite{fergus2006removing, krishnan2009fast, pan2014deblurring, pan2016blind} and blur kernel estimation strategies~\cite{xu2010two, pan2016robust, hu2016image, joshi2008psf}. In recent years, convolutional neutral networks (CNNs) have provided new approaches for image deblurring. Learning-based methods have replaced some steps in the traditional framework~\cite{sun2015learning, xiao2016learning, zhang2017learning}, and some methods learn an end-to-end mapping from blurred to latent images~\cite{tao2018scale, nah2017deep, kupyn2018deblurgan}. Compared with traditional methods, learning-based methods have advantages in computational speed, nonuniform blur restoration and robustness.

\begin{figure}[t]
\begin{center}
   \includegraphics[width=1\linewidth]{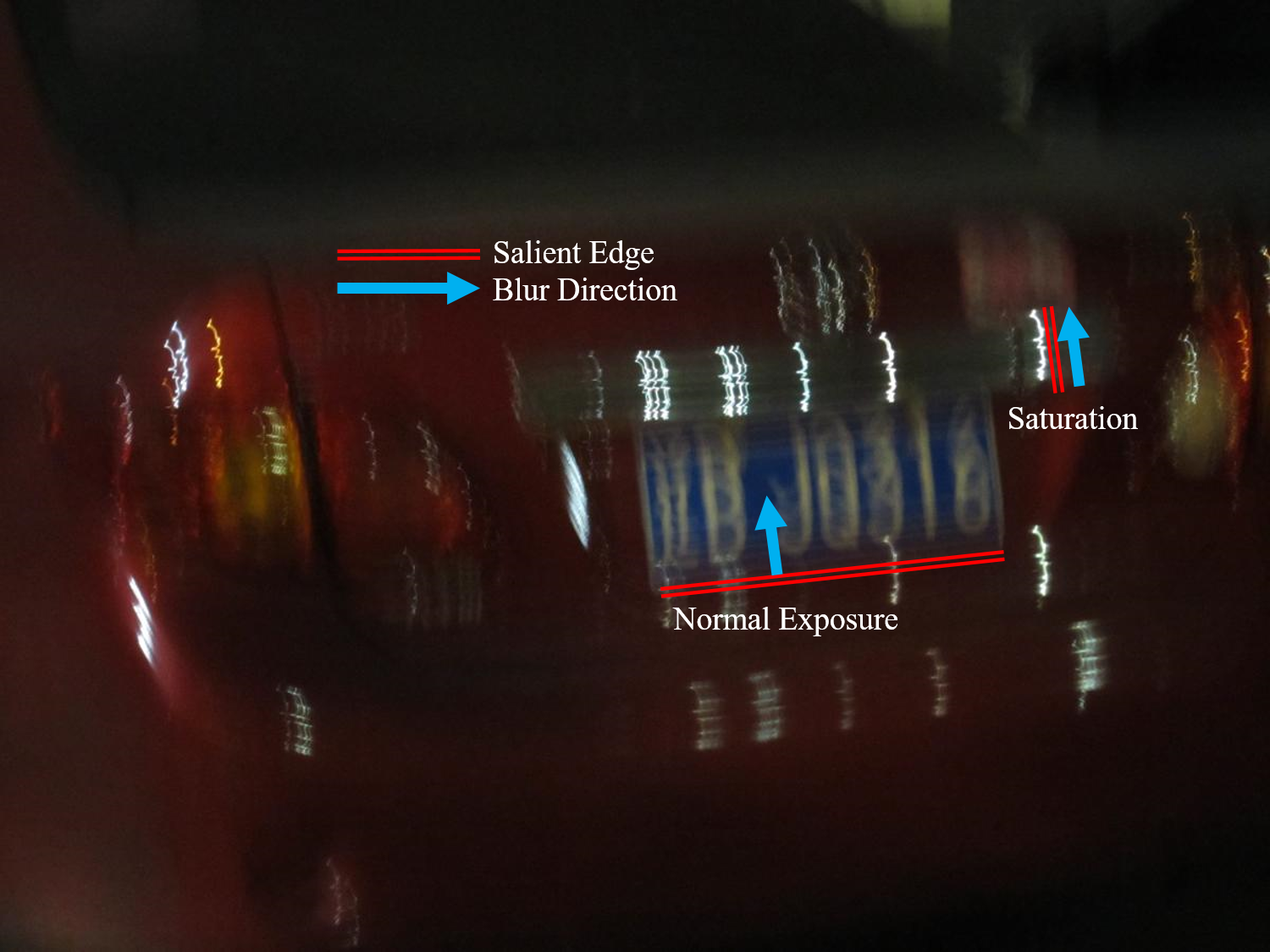}
\end{center}
   \caption{\textbf{Misleading effects caused by outliers.} The red double lines represent salient edges, while the blue arrows represent the blur direction.}
\label{fig:car}
\end{figure}

When scene illumination conditions are inappropriate, blurred images are often accompanied by overexposure or heavy noise. Regardless of whether traditional methods or learning-based methods are used, outliers such as saturated pixels or salt-and-pepper noise play negative roles in image restoration. The outliers mislead the estimation of the degenerate function~\cite{cho2011handling}. Fig.~\ref{fig:car} shows an example. Under normal circumstances, the blur information reflects the normal edge direction. However, in the example, it reflects the tangential direction of the outliers. Hence, we cannot easily assert that the edge formed by saturated pixels is a salient edge in the latent image or simply a spot without image context priors. Moreover, outliers affect the light intensity information, which leads to ringing artifacts during deblurring.

Hence, our research focuses on camera motion deblurring with outlier removal via a convolutional neural network framework. At the framework level, we select a scale-recurrent structure~\cite{tao2018scale} because it fits the “coarse-to-fine” strategy, which is effective in both learning-based and rule-based deblurring approaches. When the parameters are limited, the collection of degradation functions that a CNN can restore is finite. The scaling operation of blurred images works to increase the size of the collection of restorable degradation functions. The scale-recurrent network (SRN) can process large-scale blurred images via multiscale iteration. Moreover, the scale-recurrent structure effectively reduces the number of parameters~\cite{tao2018scale}. Within the scale approach, we design an encoder-decoder ResBlock network to achieve deblurring and a cascaded ResBlock network to perform the conversions between scales.

According to Liu \textit{et al.}~\cite{liu2013no}, the performance of image deblurring quality evaluations depends primarily on the unprocessed blur, the ringing artifacts and noise in the deblurred result. Among the above three factors, ringing artifacts most strongly affect the human visual response. Ring artifacts reflect the mid-frequency level of the deblurred image. In addition, according to Xu and Jia~\cite{xu2010two}, not all edges of an image should be treated equally because the edges whose sizes are smaller than the blur kernel are negative for blur kernel estimation. Hence, we train a salient edge selection network to supervise the deblurring. Salient edge (SE) loss is proposed to prevent the deblurred results from including ringing artifacts.

As mentioned above, outlier removal is important in camera motion deblurring. We solve this issue from the dataset. The previous learning-based methods employed two technical routes to synthesize blurred-sharp image pairs. One approach synthesizes blurred images by convolving sharp images with uniform or nonuniform blur kernels~\cite{hirsch2011fast,boracchi2012modeling}. However, this approach has difficulty in reflecting the continuous changes of blur in the image. The other approach is to generate blurred images by averaging consecutive short-exposure frames from high frame-rate videos~\cite{nah2017deep}. This approach is aimed at deblurring dynamic scenes. These CNN methods trained on dynamic scene datasets have a side effect under camera motion deblurring. They are likely to transform independent edges or gradient areas in the latent image into single strong edges.
Moreover, none of the existing datasets have considered the outlier issue. Hence, we propose a dataset synthesis approach that utilizes optical flow and saturated light streaks to more realistically simulate camera motion blur. In real camera motion cases, we prefer the dynamic objects to remain blurred rather than risk creating artifacts. (Please refer to Section 4.C and Fig.~\ref{fig:special}  for details.) The experimental results demonstrate that our dataset is effective at deblurring blurred images with outliers.

The main contributions of our work are as follows:
\begin{itemize}
    \item Based on a more accurate imaging model of the blur process, we propose a dataset synthesis approach that considers the outlier issue in camera motion deblurring. With the contribution of our dataset, the deblurring neural network is able to restore the blurred images with outliers.

    \item We propose a novel scale-recurrent network architecture for camera motion deblurring that consists of two parts: a deblurring part and an upsampling part. Both parts have distinctive capabilities and can conveniently be integrated into multiscale frameworks.

    \item We propose a salient edge loss to prevent deblurred results from ringing artifacts and design a salient edge selection network to supervise the deblurring process, which makes the deblurred results more suitable for human vision.
\end{itemize}
Based on the above contributions, our work succeeds in solving the camera motion blur issue with outliers. Through extensive empirical tests and evaluations, the proposed method is shown to outperform to the state-of-the-art methods with regard to both image restoration quality and robustness.

\section{Related Works}
Image deblurring is divided into two categories based on whether the blur kernel is known, i.e., nonblind deblurring~\cite{cho2011handling, rudin1992nonlinear, richardson1972bayesian} and blind deblurring~\cite{shan2008high, cho2011handling, whyte2014deblurring, bar2006image, fergus2006removing, krishnan2009fast, pan2014deblurring, pan2016blind, xu2010two, pan2016robust, hu2016image}. The latter is more realistic yet more ill-posed. Blind deblurring methods include both rule-based and learning-based approaches. The rule-based approaches construct frameworks utilizing Bayesian probability models and handcrafted priors. Although certain prior approaches achieve impressive effects in certain scenes, the rule-based approaches are insufficiently robust.

The rapid development of deep learning techniques has improved the ability to perform image deblurring. Previous works replaced some steps of rule-based frameworks with CNNs. For example, Sun \textit{et al.}~\cite{sun2015learning} employed a CNN to perform nonuniform kernel estimation. Zhang \textit{et al.}~\cite{zhang2017learning} proposed a deep image prior that replaced handcrafted priors and decomposed the deblurring framework into a restoration step and a denoising step. Gong \textit{et al.}~\cite{gong2017motion} utilized a CNN to estimate the motion flow during kernel estimation. More recent work has focused on creating end-to-end networks for motion deblurring. Nah \textit{et al.}~\cite{nah2017deep} learned a coarse-to-fine strategy in a rule-based deblurring framework and proposed a multiscale CNN to enlarge the receptive field for large motion cases. The proposed GoPro dataset has been widely applied in dynamic scene deblurring studies. Tao \textit{et al.}~\cite{tao2018scale} extended the multiscale CNN to a scale-recurrent network (SRN), which not only reduces the number of parameters but also increases the robustness of the deblurring effect. Gao \textit{et al.}~\cite{gao2019dynamic} further developed the SRN by adding parameter-selective sharing and nested skip connections. Kupyn \textit{et al.}~\cite{kupyn2018deblurgan} employed a generative adversarial network (GAN) for single-image motion blurring, named DeblurGAN, which provided another approach to deblurring. However, by training on the GoPro dataset~\cite{tao2018scale, gao2019dynamic, kupyn2018deblurgan, nah2017deep}, all of these methods inherited the same problem: the independent edges or the gradient areas in the latent image are likely transformed into one strong edge.

\begin{figure*}[t]
\begin{center}
   \includegraphics[width=1\linewidth]{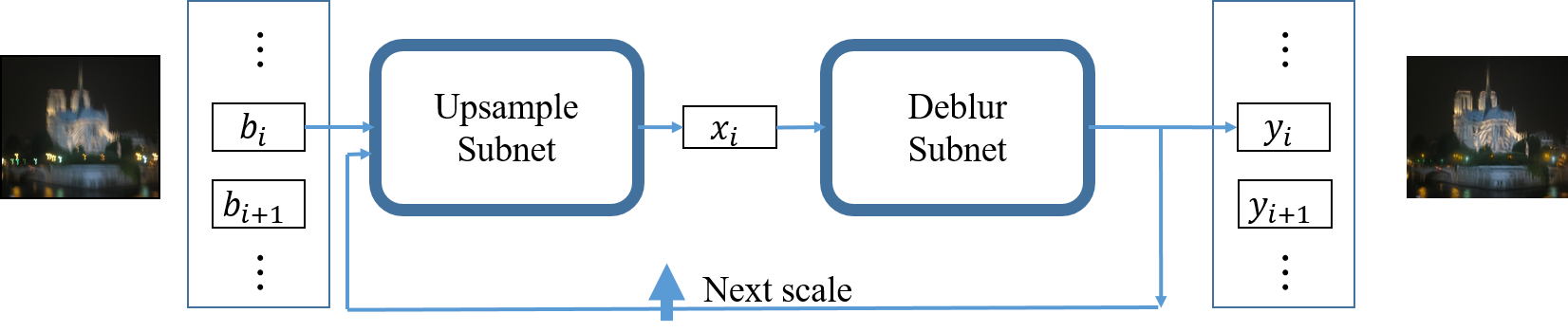}
\end{center}
   \caption{The proposed EASRN framework.}
\label{fig:framework}
\end{figure*}

\begin{figure*}[t]
\begin{center}
   \includegraphics[width=1\linewidth]{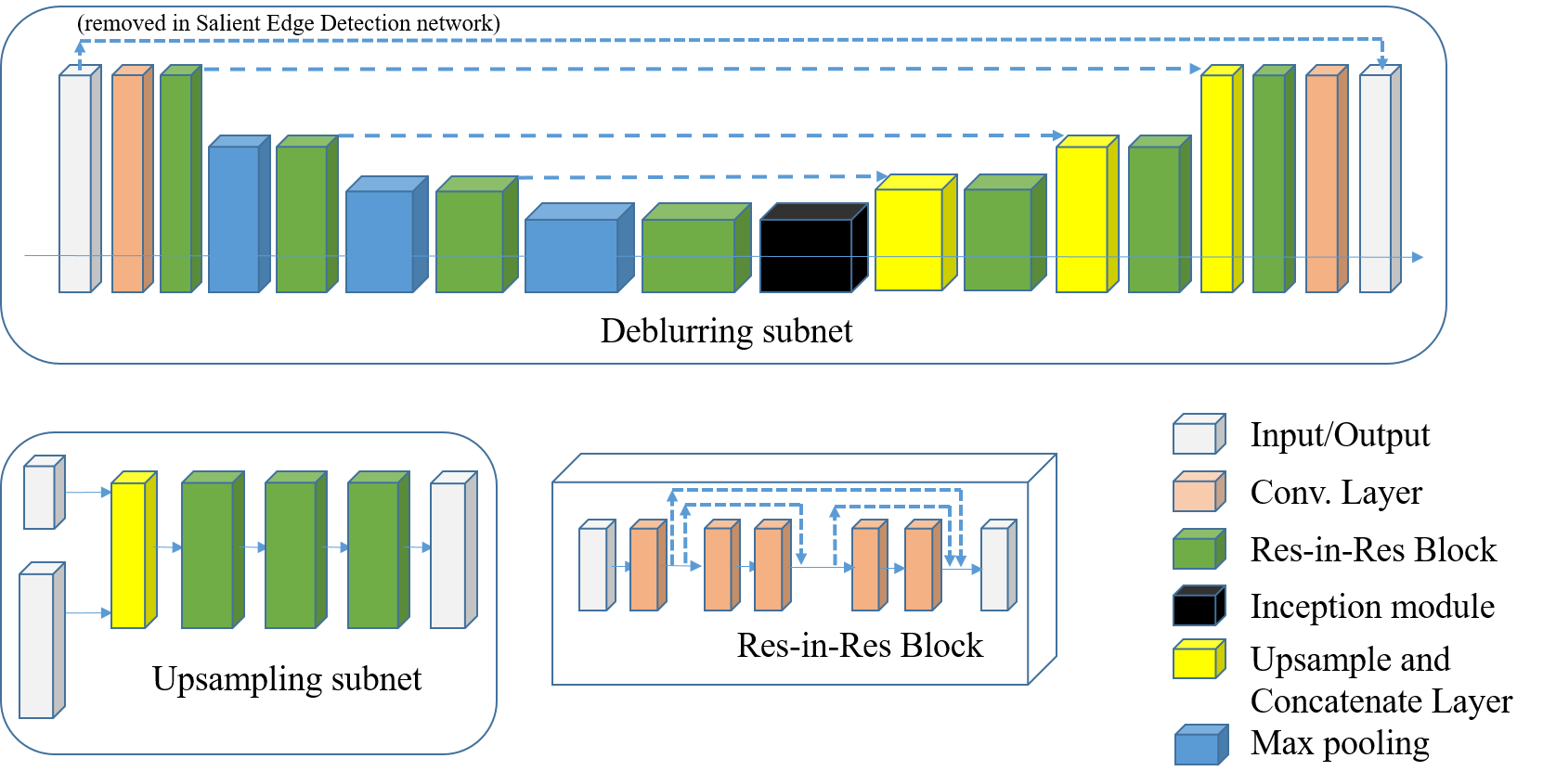}
\end{center}
   \caption{The architectures of the deblurring and upsampling subnets.}
\label{fig:network}
\end{figure*}

Regarding outlier removal in deblurring, several rule-based deblurring approaches have made contributions. Cho \textit{et al.}~\cite{cho2011handling} analyzed the effect of outliers and proposed a probability model to extend the Bayesian framework. Whyte \textit{et al.}~\cite{whyte2014deblurring} proposed a continuous probability function to replace the piecewise function in~\cite{cho2011handling} to achieve a better deblurring effect. Pan \textit{et al.}~\cite{pan2016robust} analyzed how outliers mislead algorithms in blur kernel estimation and proposed a confidence function to remove the outliers during kernel estimation, which enabled the algorithm to address blurred images with outliers. Dong \textit{et al.}~\cite{dong2017blind} proposed a data fidelity term in the loss function to replace the confidence function in the approach of Pan \textit{et al.}~\cite{pan2016robust}. Gong \textit{et al.}~\cite{gong2017motion} proposed to sequentially identify inliers and gradually incorporate them into the kernel estimation process. These three methods~\cite{pan2016robust, gong2017motion, dong2017blind} belong to the same technical route and have the same ringing artifact issue. Hu \textit{et al.}~\cite{hu2014deblurring} attempted to employ light streaks to estimate the blur kernel, and the proposed method performs impressively in various situations. However, when the scene has no obvious light streaks, this method is unsatisfactory. To the best of our knowledge, no existing learning-based approach is targeted toward deblurring images with outliers.

\section{Proposed Method}

In this section, we describe our model development. Fig.~\ref{fig:framework} provides an overview of the proposed network architecture, called the edge-aware scale-recurrent network (EASRN). The network has a recurrent architecture consisting of two parts: a deblurring subnet and an upsampling subnet. Fig.~\ref{fig:network} shows the details of the deblurring and upsampling subnets. A blurred image $B$ is first decomposed into an $N$-scale Gaussian pyramid $\{b_i\}, i=1,2,...,N$. We deblur $\{b_i\}$ from a minimum scale to a maximum scale. At scale $i$, the restoration process is as follows:
\begin{equation}
    y_i=Deblur(x_i),
\end{equation}
\begin{equation}
    x_{i+1}=Upsample(y_i, b_{i+1}),
\end{equation}
where $Deblur$ denotes the deblurring subnet as a function, and $Upsample$ denotes the upsampling subnet as a function. The intermediate variable $x_i$ represents the preprocessed blurred image, which is deblurred by the deblurring subnet to obtain the latent image $y_i$. Then, $y_i$ is upsampled and concatenated with the next-scale blurred image $b_{i+1}$, forming the input to the upsampling subnet to obtain the next-scale intermediate variable $x_{i+1}$. The latent image at the full resolution, $y_N$, is the final output. At the first scale, the upsampling subnet is omitted. With deblurring from coarse to fine, our model can deal with larger motion blur.

\subsection{Deblurring Subnet}
The deblurring subnet is designed to deblur an input image. Our deblurring subnet employs the encoder-decoder network structure. The encoder outputs a feature map that represents the input. The decoder uses the feature maps from the encoder and tries to achieve the closest match to the intended output. The encoder uses max pooling to downsample the feature maps, while the decoder uses deconvolution upsampling. Skip connections are added between corresponding feature maps in the encoder-decoder to combine different information levels.

Here, we introduce residual-in-residual blocks (Res-in-Res Block) in both the encoder and decoder parts. Each Res-in-Res Block contains two ResBlocks~\cite{he2016deep} in a residual unit with a short skip connection. At the same time, an inception module~\cite{SzegedyGoing} is adopted between the encoder and decoder to capture multiscale information. The four filter sizes in the inception module are set to 1, 3, 5, and 7 separately and the channel numbers of the encoder-decoder pairs are 32, 64, 128 and 256. Two additional convolution layers are employed at the front and back ends to extract features from the input and combine them into the output. A residual connection from the input blurred image to the output deblurred result is added to improve residual learning. Except for the inception module, all kernel sizes are set to 3. A leaky rectified linear unit (LReLU)~\cite{maas2013rectifier} is employed as the activation function for the entire subnet. Compared to Tao \textit{et al.}~\cite{tao2018scale} who used large convolution kernel sizes and cascaded ResBlocks to form encoder/decoder blocks, our structure has fewer parameters and better performance.

\subsection{Upsampling Subnet}
The upsampling subnet is designed to fuse the lower-scale output of the deblurring subnet with the next-highest scale of the blurred image. The ideal situation is to acquire the low-frequency information from the lower-scale output and the high-frequency information from the upper-scale blurred image. During the process of scale transformation, Tao \textit{et al.}~\cite{tao2018scale} and Gao \textit{et al.}~\cite{gao2019dynamic} directly upsampled the lower-scale intermediate latent image and concatenated it with the upper-scale blur image to form the input to the next scale. This design requires the first encoder block to take on an additional fusion task. We separated this task and designed an extraction-and-upsampling subnet to update the multiscale fusion effect.

The upsampling subnet consists of an upsampling and concatenation layer and 3 Res-in-Res Blocks. The Res-in-Res Blocks have 32 channels. The number of channels in the output is the same as that of the input. Similar to the deblurring subnet, we adopt $3\times 3$ filters and the LReLU. Using this subnet, the immediate deblurring result will restore more high-frequency textures.

\subsection{Training Losses}
As mentioned above, the image quality of the deblurred result reflects three main aspects: noise level, ringing artifacts and residual blur. The outliers in blurred images increase the difficulty of solving the above three problems. Hence, we design a loss for each aspect.

For noise reduction and data fidelity, we use the $L_1$ loss for each scale. The ground truth $G$ is decomposed to obtain an $N$-scale Gaussian pyramid $\{g_i\}, i=1,2,...,N$, and the fidelity loss is calculated as follows:
\begin{equation}
    \mathcal{L}_f=\frac{1}{N}\sum^N_i{{\Vert y_i-g_i \Vert}_1},
\end{equation}
Compared with $L_2$ loss, $L_1$ loss has better noise tolerance and performs better for ringing artifact removal.

To remove ringing artifacts, we employ the $L_1$ distance between the salient edge maps of the deblurred result and ground truth to represent the ringing artifact loss. A salient edge-detection network (SEDNet) is designed to obtain the salient edge map. An example salient edge-detection case is shown in Fig.~\ref{fig:museum} and discussed in detail in Section 4.B. 

SEDNet has a similar structure to the deblurring subnet, except that the residual connection from the input to the output is removed since it is not an identical mapping. This similar structure makes the receptive fields of the two networks match. Only the full resolution deblurred result is required to have a salient edge map that is close to the ground truth because it is uncertain whether the ringing artifacts of intermediate latent images are incurable. If the salient edges at each scale are used for supervision during training, the deblurred result will lead to oversmoothing and loss of small details. 

Thus, the ringing artifact loss (or SED loss) is represented as follows:
\begin{equation}
    \mathcal{L}_s=w_s {\Vert SED(y_N)-SED(g_N)\Vert}_1, 
\end{equation}
where $SED$ denotes SEDNet as a function and $w_s$ represents a weight coefficient.

The residual blur of the deblurred result reflects the loss of detail. We employ perceptual loss~\cite{JohnsonPerceptual} at each scale to enhance the detail in the deblurred results. The deblurred result and the ground truth are both fed into the pretrained VGG-19 network, and  perceptual loss is the difference between their corresponding feature maps. The front layers of the VGG network characterize the textual information or the low-level features of an image, therefore, we select $conv_{1,2}$ and $conv_{2,2}$ from VGG-19 as the feature maps to be compared. However, the deblurred result will be gridded if only perceptual loss is used. Hence, we also employ total variation regularization to suppress the grid effect. The detailed enhancement loss is denoted as
\begin{equation}
    \mathcal{L}_v=\sum^N_i{(w_p \sum_j{\Vert vgg_j(y_i)-vgg_j(g_i) \Vert}^2_2+w_t {\Vert \nabla y_i \Vert}^2_2)},
\end{equation}
where $vgg_j(\cdot)$ represents feature map $j$ of the VGG-19 network, $j=1,2$ or $ 2,2$, and $w_p$ and $w_t$ represent the corresponding weight coefficients.

As a result, we restrict the output from three aspects: noise level, ringing artifact removal and detail enhancement. The total loss is represented as follows:
\begin{equation}
    \mathcal{L}=\mathcal{L}_f+\mathcal{L}_s+\mathcal{L}_v.
\end{equation}

\subsection{Dataset}

\begin{figure*}[t]
\begin{center}
   \includegraphics[width=0.9\linewidth]{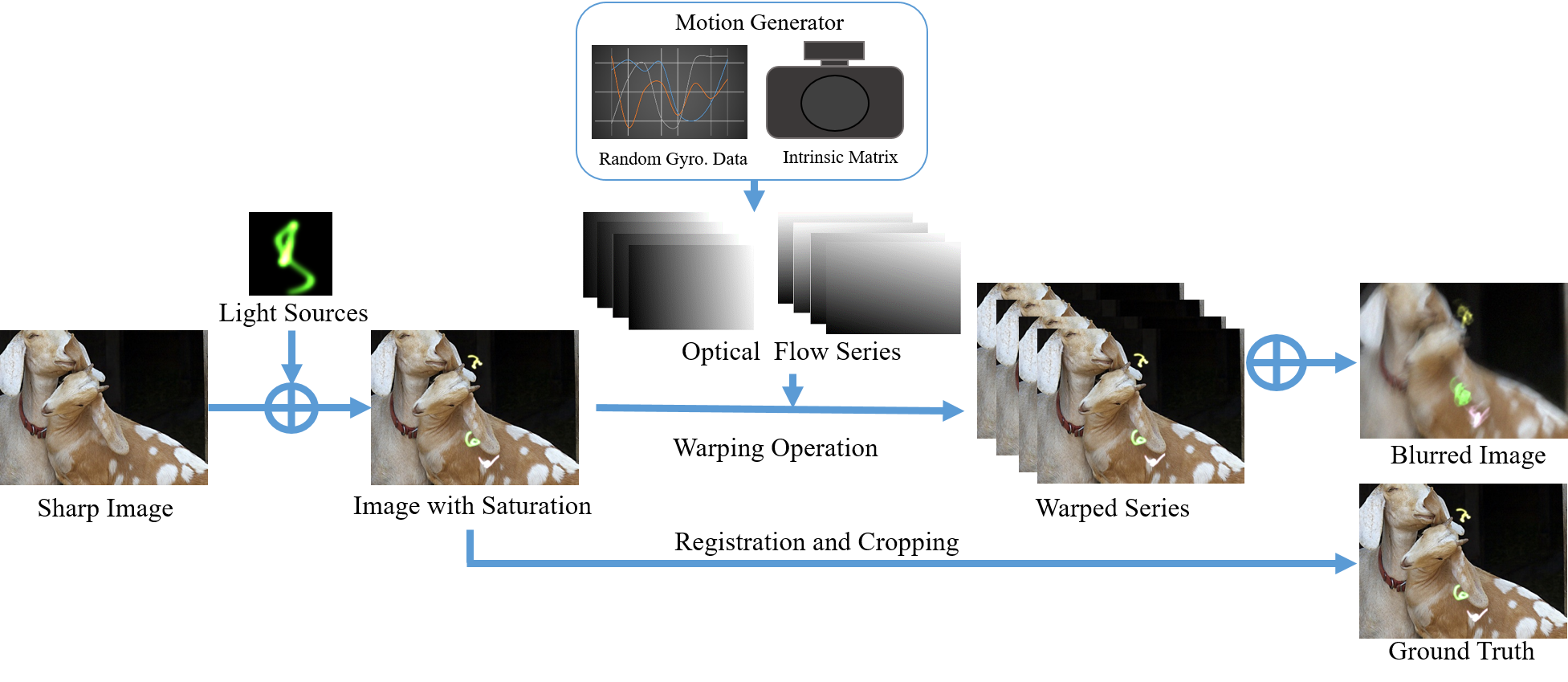}
\end{center}
   \caption{Flowchart for synthesizing blurred/sharp image pairs.}
\label{fig:dataset}
\end{figure*}

The methods to acquire blurred-sharp image pairs directly from hardware devices are difficult. A typical approach for obtaining synthetic image pairs is to generate blurred images by averaging consecutive short-exposure frames from high-frame-rate videos. However, the GoPro dataset produced using this approach is not appropriate for our task because the scenes are dynamic; in contrast, we aim to restore images that suffer from camera motion blur. Moreover, this approach cannot simulate real overexposure outliers since the dynamic range of each frame is cut off. Another widely adopted method is to create blurred images by convolving the sharp image with either uniform or nonuniform blur kernels. The fidelity of the blur kernels determines the value of the synthetic dataset. Boracchi and Foi~\cite{boracchi2012modeling} analyzed the coordinate function of each pixel in image space under real-world camera motion conditions. We followed this approach and assumed that the camera motion is a Markovian process. Next, we calculated the coordinates of each pixel to obtain the optical flow map at time $t$. Then, we warp the sharp image utilizing the optical flow map and subpixel interpolation to obtain the intermediate image at time $t$. Finally, we accumulate and average the intermediate image series to obtain the blurred image and add Gaussian noise with a standard deviation of $\sigma$ after the blurring process is complete. Compared with previous approaches, our synthesis method is closer to real camera motion situations.

\begin{figure}[t]
\begin{center}
   \includegraphics[width=1\linewidth]{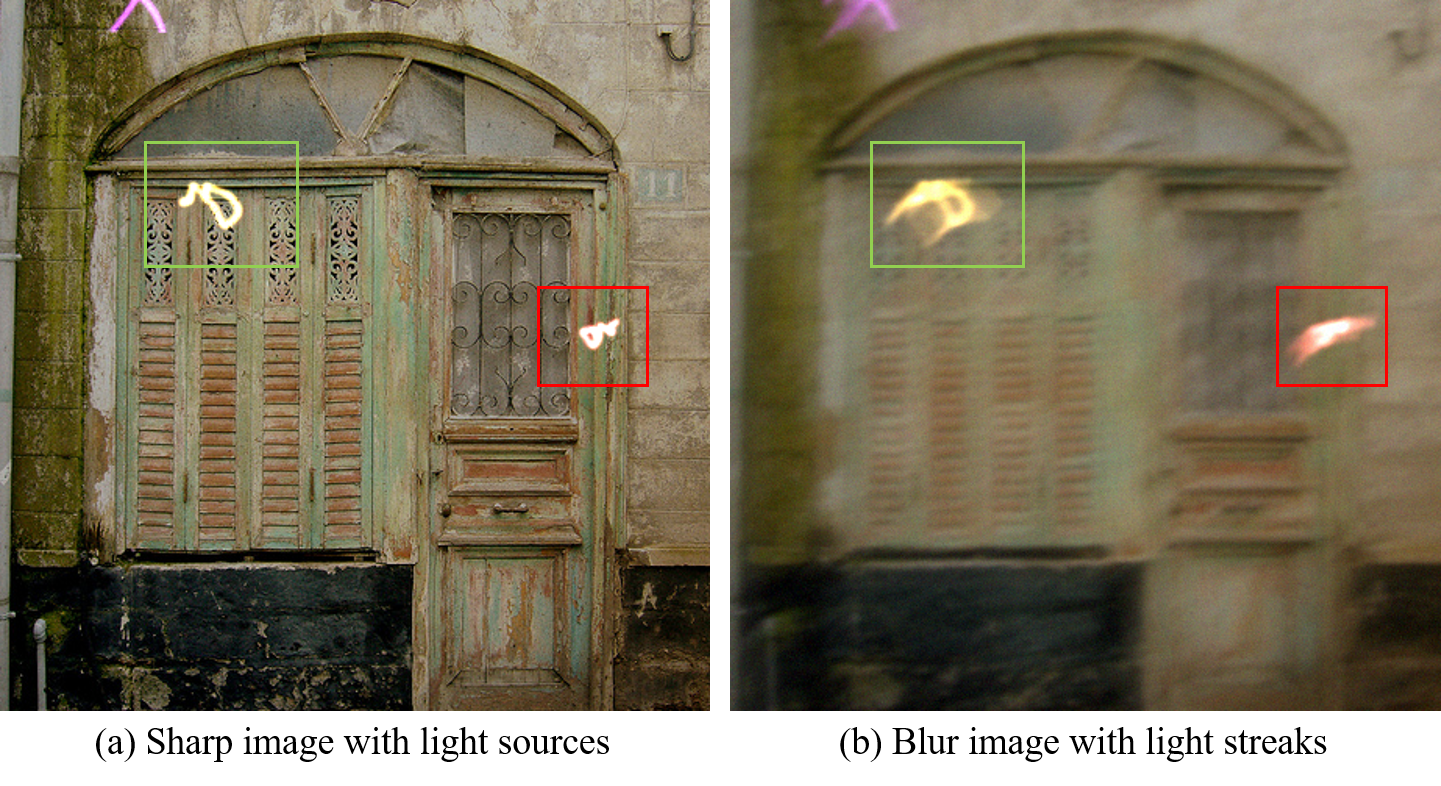}
\end{center}
   \caption{An example of a synthesized blurred/sharp pair.}
\label{fig:OEcase}
\end{figure}

Overexposure light sources may create light streaks in blurred images. Hu \textit{et al.}~\cite{hu2014deblurring} indicates that distinguishing the light streaks in blur images and the other salient edges is advantageous for kernel estimation. We expect the CNN-based model to learn to extract and utilize the blur information in the light streaks—or at least to distinguish them from textures of normal exposure. To simulate the degradation feature of outliers in blur situations, we apply an extra ‘print light sources’ processing step to part of the synthetic dataset. First, due to the variety of light sources in the real scenes, we generate light sources with random shapes using the approach of random trajectory generation proposed by Boracchi and Foi~\cite{boracchi2012modeling}. The intensity of the light sources is 1 to 10 times that of the maximum dynamic range of the sharp image, simulating the cutoff effect of the limited dynamic range. The intensities of the three channels are treated independently to simulate colorful light sources. Next, we randomly select a certain percentage of the sharp images before blur processing and insert random numbers of the generated light sources into them. Then, the sharpened images with light sources are blurred by the proposed blur synthesis method. Finally, the values of both the sharp and blurred images are clipped to match the original dynamic range. Because blur synthesis processing leads to a center offset of the generated image pair, the sharp image is registered correspondingly to obtain the ground truth. A flowchart of this process is shown in Fig.~\ref{fig:dataset}, and Fig.~\ref{fig:OEcase} shows an example blurred/sharp pair with the extra ‘print light sources’ processing.

\subsection{Implementation Details}
We implemented all our models using the TensorFlow platform. Training was performed on two NVIDIA TITAN xp GPUs.

\textbf{Data Preparation} 
For EASRN, we excluded the blurred images and selected 3,155 high-resolution images from the Flickr2K and DIV2k datasets as the sharp images. After random cropping, flipping and gamma correction operations, we obtained 50,480 sharp images with a size of $512\times 512$. On one-third of the sharp images, we added 2 to 20 light streaks randomly. The maximum relative shift in optical flows is constrained to 30 pixels by setting the appropriate range of the camera motion blurring. The standard deviation of Gaussian noise $\sigma$ is set to no more than 0.02. After obtaining the blurred/sharp image pairs, we divided the dataset into 46,000 pairs for training and 4,480 pairs for evaluation.

For SEDNet, we employed the augmented BSDS data from Xie and Tu~\cite{xie2015holistically} as our dataset. The BSDS 500 dataset provides an empirical basis for research on image segmentation and boundary detection, and it includes 500 hand-labeled image-mask pairs. After scaling, rotation and flipping operations, the augmented BSDS 500 dataset includes 28,801 pairs.

\textbf{Model Parameters}
The blurred image and ground truth are decomposed to 3 scales; thus, $N=3$  during training. The deblurring processes move from the smallest to the largest scale. The input image of the minimum scale is set as the corresponding scale of the blurred image, $x_1=b_1$. The parameters to control the weights of different losses were set as follows: $w_p=3\times 10^{-6}$, $w_t=0.8$ and $w_s=2.4$.

\textbf{Training Details}
For EASRN, we employed the Adam optimizer and set $\beta_1=0.9$ and $\beta_2=0.999$ for training. The learning rate decayed exponentially every 40 epochs from its initial value of $1\times 10^{-4}$ at a power of 0.1. In each iteration, we sampled a batch of 8 blurred/sharp image pairs and randomly cropped them to $256\times 256$-pixels for both the training input and the ground truth. Then the pretrained VGG-19 and SEDNet were directly loaded as fixed networks. The complete training process requires approximately 48 hours for 80 epochs.

For SEDNet, the Adam optimizer and the learning rate were set to the same values used for EASRN. $L_1$ loss was employed to train the SEDNet. In each iteration, we sampled a batch of 8 image-mask pairs and randomly cropped them to $64\times 64$ pixels to form the training input and the ground truth. After approximately 85 epochs, the loss curve begins to oscillate due to dataset size limitations. The SEDNet training requires approximately 4 hours to complete. It is pretrained before the deblurring network and fixed when training EASRN.

\section{Experimental Results}

In this section, we analyze the effectiveness of the proposed contributions by controlled experiments and intermediate results analysis. Section 4.A explains why the separated deblurring and upsampling subnets benefit the image restoration. Section 4.B indicates the positive role of SED loss for ringing artifact reduction. Section 4.C discusses the application scope of the proposed dataset and its advantages for camera motion blur removal. Section 4.D shows how much the three contributions of our work affect outlier removal in deblurring by an ablation study. 

Furthermore, we compare the proposed method with both other state-of-the-art learning-based deblurring methods and some traditional deblurring methods. Section 4.E reports the quantitative indicators on typical test sets. In this part, we focus on the comprehensive performance of the algorithms, rather than the advantages and disadvantages caused by the differences between their components.

We selected 3 credible datasets to evaluate the performances of our method and those of the state-of-the-art methods. Köhler et al.’s~\cite{kohler2012recording} dataset, which consists of 48 images with 12 different kernels, is a standard benchmark dataset for evaluating camera motion deblurring approaches. This dataset was created by recording real camera motion from videos that captured the 6-degrees-of-freedom motion trajectory during camera was imaging. The nonuniform blurred images are generated patch by patch. The GoPro dataset~\cite{nah2017deep} is formed from 240 frames-per-second video sequences acquired by a GoPro Hero 4 camera. The blurred images are generated by averaging consecutive short-exposure frames. The GoPro dataset~\cite{nah2017deep} is also commonly used as a benchmark for evaluating dynamic scene deblurring approaches. Lai \textit{et al.}~\cite{lai2016comparative} performed a comparative study of rule-based single-image blind deblurring approaches. Lai’s dataset~\cite{lai2016comparative} consists of three parts: a uniform-blur part, a nonuniform-blur part, and a real-world blur part. The uniform-blur and nonuniform-blur parts are generated by a synthesis method similar to that used to create Köhler \textit{et al.’s}~\cite{kohler2012recording} dataset. The real-world part consists of 100 real blurred images published in previous deblurring works. The generated types of evaluation datasets prefer deblurring models whose training datasets were generated in similar ways. Hence, we combine both subjective evaluation indexes and objective evaluation indexes to enable a fair and comprehensive comparison. 

\subsection{Effectiveness of the Upsampling Subnet}

\begin{figure}[t]
\begin{center}
   \includegraphics[width=1\linewidth]{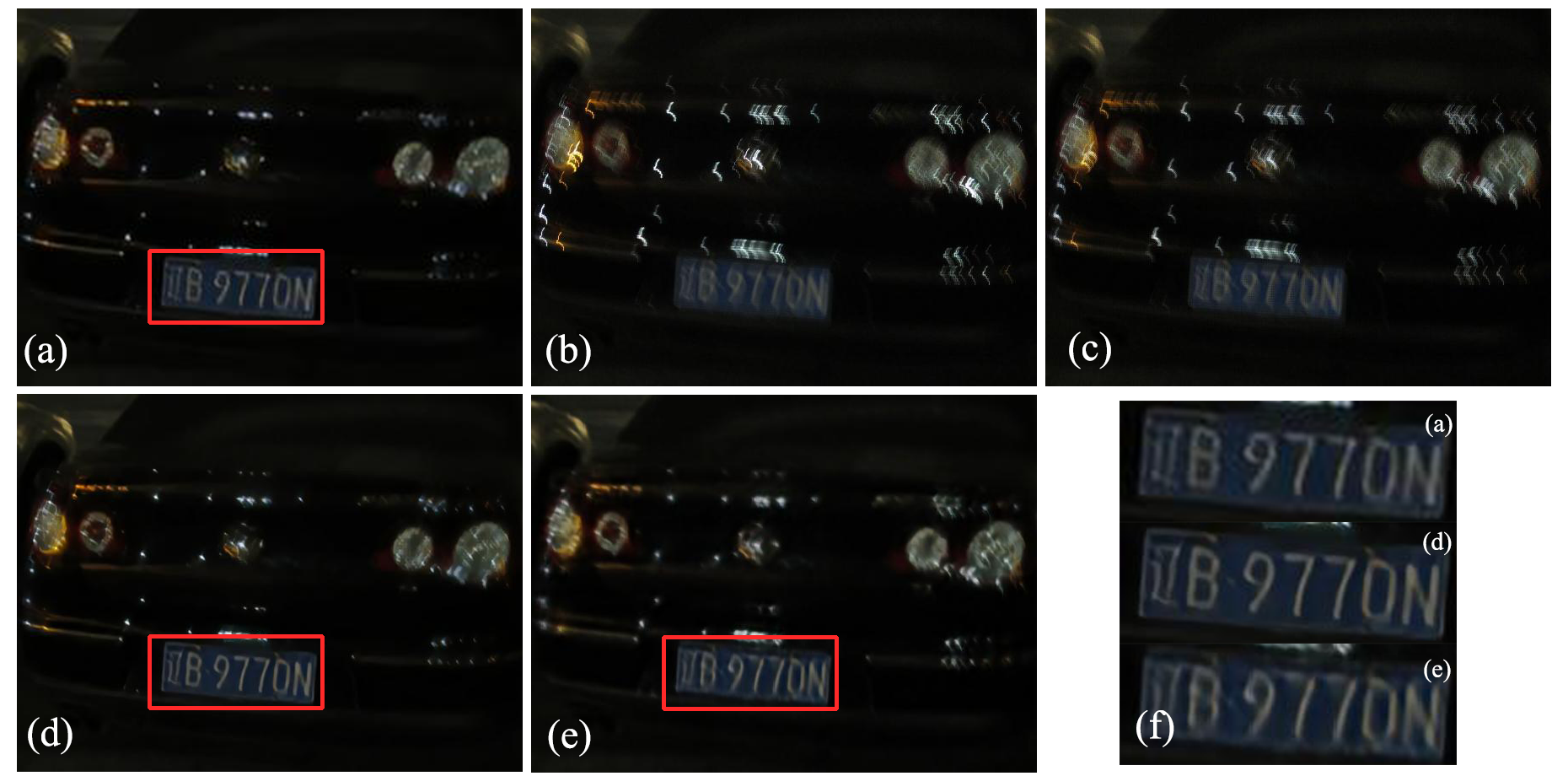}
\end{center}
   \caption{\textbf{Details of the intermediate results.}(a) the deblurring output from the lower scale after the upsampling operation; (b) the blurred input in the current scale; (c) the output of the upsampling subnet in the current scale; (d) the deblurring output in the current scale; (e) the deblurring result from the model without an upsampling subnet; (f) detail enlargements of (a), (d) and (e).}
\label{fig:interm}
\end{figure}

To demonstrate the effectiveness of the separated upsampling subnet, we ablated the upsampling subnet of EASRN and trained the modified network. In this model, the latent output $y_{i-1}$ from the lower scale is upsampled and concatenated directly with the current-scale blurred image $b_i$ to form the input for the current scale. Fig.~\ref{fig:interm} shows the details of the intermediate results from the proposed method and the model without the upsampling subnet.  Fig.~\ref{fig:interm}(a) shows the deblurred output from the lower scale after the upsampling operation. Fig.~\ref{fig:interm}(b) shows the blurred input at the current scale. Fig.~\ref{fig:interm}(c) shows the output of the upsampling subnet at the current scale. Fig.~\ref{fig:interm}(d) shows the deblurred output at the current scale. Fig.~\ref{fig:interm}(e) shows the deblurred result of the model without the upsampling subnet, and Fig.~\ref{fig:interm}(f) shows detail enlargements of (a), (d) and (e). The output of the upsampling subnet fuses the low-frequency information of $y_{i-1}$ with the high-frequency information of $b_i$. The deblurred result in Fig.~\ref{fig:interm}(d) has more texture and higher resolution than the image in Fig.~\ref{fig:interm}(a). However, the network without upsampling subnet generates a worse deblurred result, as shown in Fig.~\ref{fig:interm}(e), which results in structural distortion and blur residuals. Moreover, with the preprocessing from the upsampling subnet, the deblur subnet is independent on every scale. The weight sharing structure for each scale takes into account the information of each frequency. The upsampling subnet is beneficial in making the deblurring subnet extract features more fully.

\subsection{Effectiveness of SED Loss}

\begin{figure}[t]
\begin{center}
   \includegraphics[width=1\linewidth]{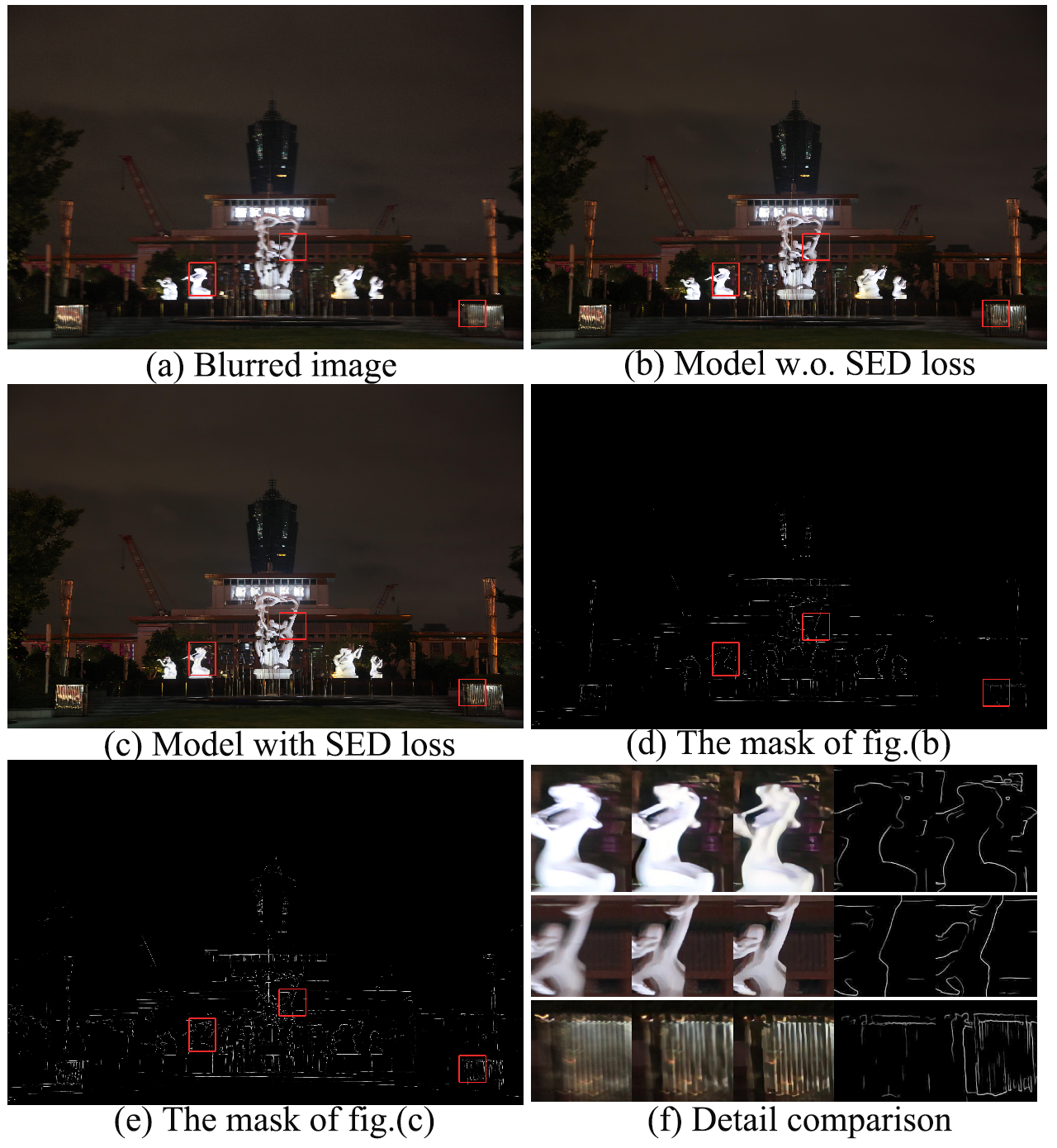}
\end{center}
   \caption{A comparison between models with and without SED loss.}
\label{fig:museum}
\end{figure}

To demonstrate the effectiveness of SED loss, we trained the proposed network only with $\mathcal{L}=\mathcal{L}_f+\mathcal{L}_v$. The obtained model is named EASRN w.o. SED. To indicate the role of SED loss more intuitively, the image deblurring case shown in Fig.~\ref{fig:museum} is employed to illustrate the effect: Fig.~\ref{fig:museum}(a) is the blurred image; Fig.~\ref{fig:museum}(b) is the output from the model without SED loss; Fig.~\ref{fig:museum}(c) is the output from the model with SED loss; Fig.~\ref{fig:museum}(d) is the salient edge mask of Fig.~\ref{fig:museum}(b) from SEDNet. Fig.~\ref{fig:museum}(e) is the mask of Fig.~\ref{fig:museum}(c); and Fig.~\ref{fig:museum}(f) shows the detail enlargements of the corresponding red rectangles from the above figures. Distortions of the edge spatial intensity distribution result in ringing artifacts. The SED loss applies additional constraints to the pixels whose corresponding regions in the ground truth are salient edges. The SED loss requires that the salient edges appear in the deblurred results in terms of position and contrast. Hence, the deblurred results from EASRN are better at ringing artifact removal than the model without SED loss.

\subsection{Characteristic of the Proposed Dataset}

\begin{figure}[t]
\begin{center}
   \includegraphics[width=1\linewidth]{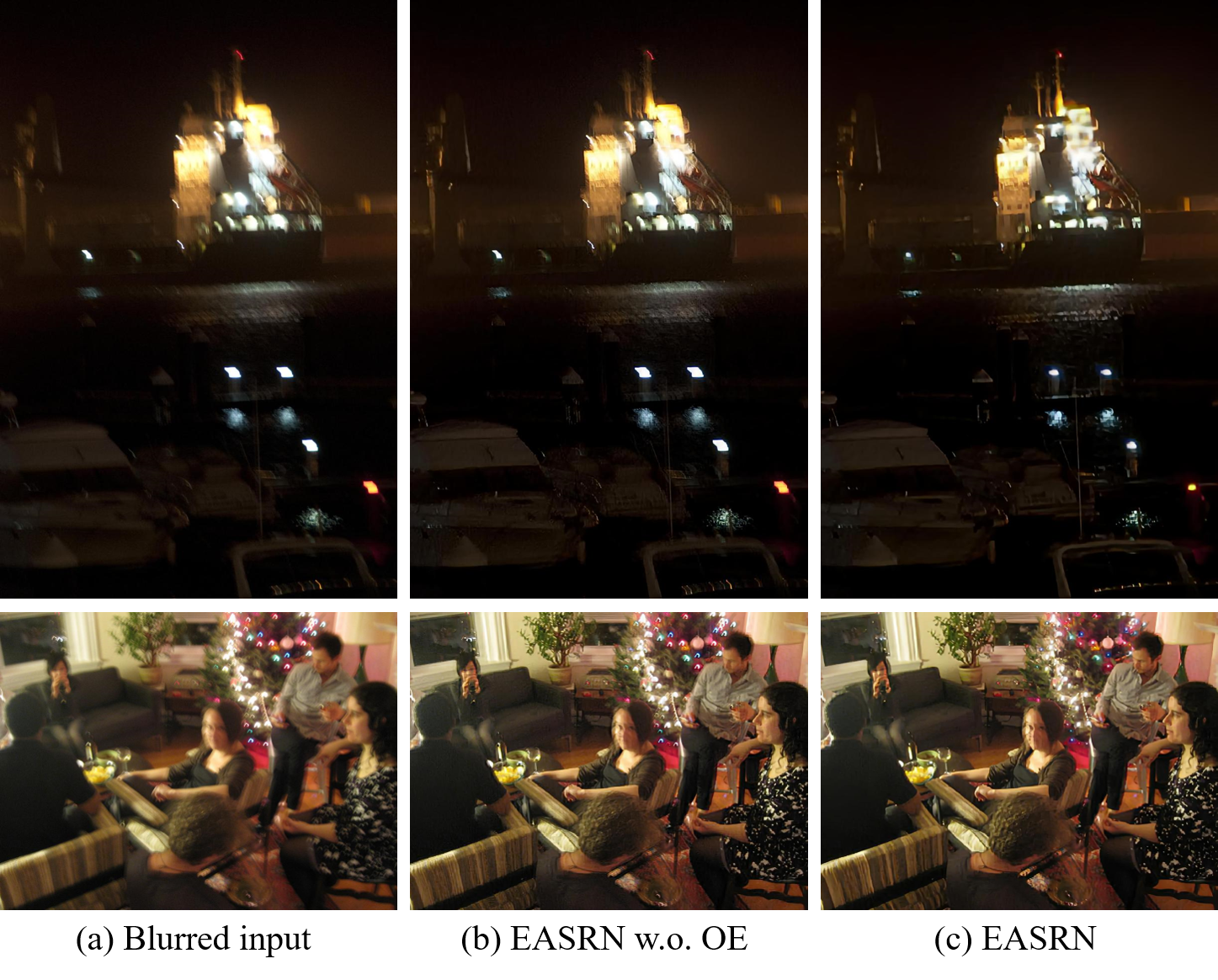}
\end{center}
   \caption{A comparison on different training datasets.}
\label{fig:outlier}
\end{figure}

Our dataset is proposed to simulate the camera motion blur in photography. The blur in the dynamic scene is not our restorative target; instead, the intent of the proposed dataset is that the blur caused by a limited depth of field and dynamic scenes should remain unchanged after deblurring. Considering this practicality, we set the size of blur kernel to no more than $64\times 64$ pixels for a $256\times 256$ image.

We performed an ablation study on the proposed dataset to prove the effectiveness of the outlier handling. In addition to the proposed synthetic dataset, we also trained our model on the dataset without adding light streaks, and obtained another version named EASRN w.o. OE. The experimental results are shown in Fig.~\ref{fig:outlier}. EASRN w.o. OE cannot handle the overexposure outliers, which results in blur residuals. As a contrast, EASRN can remove the motion blur caused by the outliers while obtaining clearer edges and richer details.

\begin{figure}[t]
\begin{center}
   \includegraphics[width=1\linewidth]{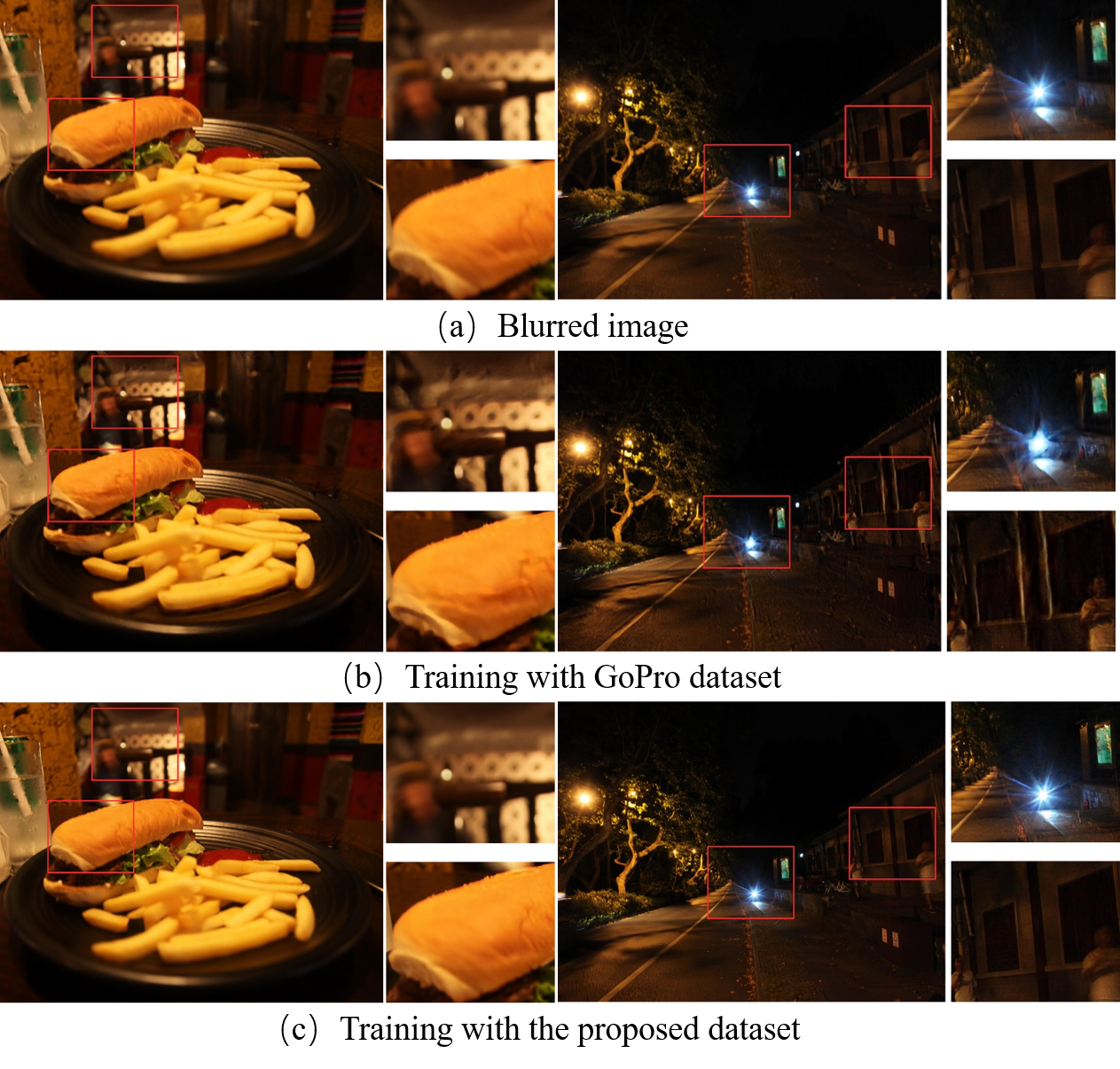}
\end{center}
   \caption{\textbf{Special deblurring cases.} From left to right are “Burger”, a blurred image with blur background and “Night view”, a blurred image with dynamic scene.}
\label{fig:special}
\end{figure}

To demonstrate the difference between the proposed dataset generation and the approach that averages consecutive short-exposure frames from high-frame-rate videos, we show some special deblurring cases in Fig.~\ref{fig:special}. “Burger” is a blurred image with a blurred background (out of focus), and “Night view” is a blurred image with a dynamic scene. In both pictures, there are blurs that the photographer wants to keep. Only camera motion blur is required to be removed. Compared to forcibly sharpening edges that should not be sharpened, we tend to retain these blurs. In contrast to GoPro dataset~\cite{nah2017deep}, the achieves results with more natural visual effects, as shown in Fig.~\ref{fig:special}. In addition, simulating the overexposure regions correctly is difficult to accomplish with the short-exposure-frame-averaging approach, which makes it difficult for the deblurring model to remove the outliers.

However, the additional ‘print light streaks’ processing may introduce side effect to the other deblurring results, especially when the proportion of light streaks added is too high. The quantitative comparison results on Kohler’s dataset~\cite{kohler2012recording} are listed in Table~\ref{table:kohler}. The PSNR and SSIM of EASRN are lower than those of EASRN w.o. OE because Kohler’s dataset does not include blurred images with outliers. This result indicates that the model has gained the ability to handle outliers by slightly sacrificing the image restoration quality. We balance the performance by adding 1/3 overexposed images to the synthetic dataset.

\subsection{Ablation study}
To demonstrate the effect of our proposed dataset, loss and architecture, we employ the SRN model~\cite{tao2018scale} for comparison. All three elements have two options, i.e., SRN or EASRN, $L_2$ loss or our loss and GoPro dataset~\cite{nah2017deep} or our dataset. The training parameters of all the tested models are the same. We employ the 20 saturated blurred images in the synthetic portion of Lai’s dataset~\cite{lai2016comparative} for evaluation. The quantitative results are listed in Table~\ref{table:ablation}. Among the models, EASRN with our loss and our dataset performs the best. Among the three elements, the dataset makes the greatest contribution to increasing the deblurring quality. The effects of the architecture and SED loss are mainly reflected in visualization.


\begin{table}[htbp]
\renewcommand{\arraystretch}{1.3}
\renewcommand\tabcolsep{12.0pt} 
\caption{Ablation study on the synthetic portion of Lai's dataset (saturated images).}
\label{table:ablation}
\begin{center}
\begin{tabular}{c|c|c|c|c}
\hline
Architecture &  Dataset & Loss & PSNR & SSIM\\
\hline\hline
\multirow{4}{*}{SRN} & \multirow{2}{*}{GoPro} & $L_2$ & 20.61 & 0.6341 \\
\cline{3-5}
~ & ~ & Ours & 20.09 & 0.6103 \\
\cline{2-5}
~ & \multirow{2}{*}{Ours} & $L_2$ & 21.70 & 0.7214\\
\cline{3-5}
~ & ~ & Ours & 22.04 & 0.7259 \\
\hline
\multirow{4}{*}{EASRN} & \multirow{2}{*}{GoPro} & $L_2$ & 20.59 & 0.6351 \\
\cline{3-5}
~ & ~ & Ours & 19.90 & 0.6053 \\
\cline{2-5}
~ & \multirow{2}{*}{Ours} & $L_2$ & \textcolor[rgb]{0,0,1}{22.19} & \textcolor[rgb]{0,0,1}{0.7287}\\
\cline{3-5}
~ & ~ & Ours & \textcolor[rgb]{1,0,0}{22.59} & \textcolor[rgb]{1,0,0}{0.7425} \\
\hline
\end{tabular}
\end{center}
\end{table}

\subsection{Comparisons with outlier handling methods}
\begin{figure*}[t]
\begin{center}
   \includegraphics[width=1\linewidth]{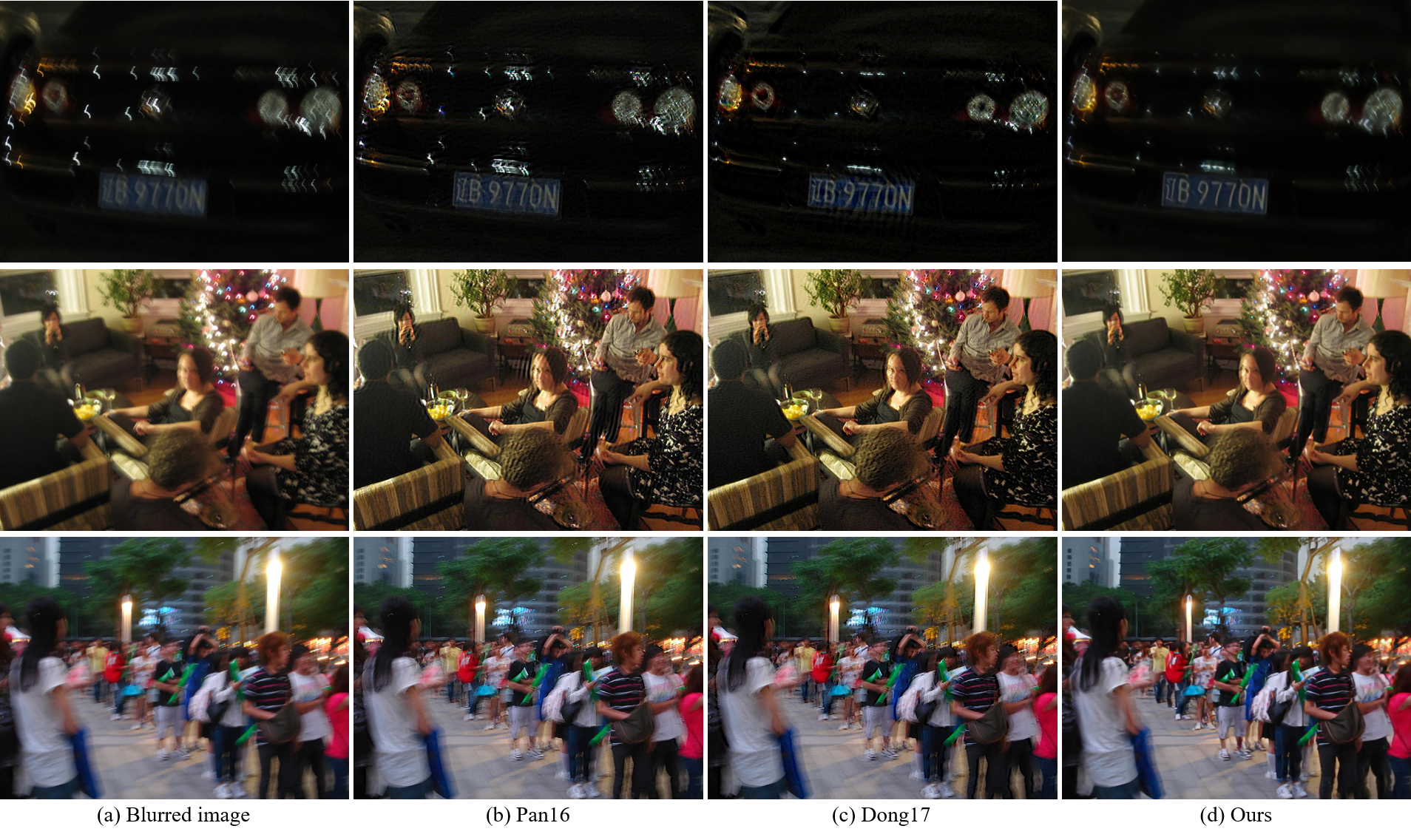}
\end{center}
   \caption{A comparison of deblurring methods on the real saturated images.}
\label{fig:saturated}
\end{figure*}

We compare our method with the state-of-the-art deblurring methods which are designed for the outlier handling, including Pan \textit{et al.}'s method (Pan16)~\cite{pan2016robust} and Dong \textit{et al.}'s method (Dong17)~\cite{dong2017blind}. In addition, to verify the performance in the real scenes, we select saturated blurred images from the real part of Lai's dataset~\cite{lai2016comparative} for evaluation. The comparison results are shown in Fig.~\ref{fig:saturated}. Pan16~\cite{pan2016robust} cannot effectively suppress the negative effects of the saturated pixels in deblurring process, resulting in blur residuals (as shown in the 'car' and 'street' images) or strong ring artifacts around the edges (as shown in the 'family' image). Although Dong17~\cite{dong2017blind} can removal light streaks caused by the outliers, some ring artifacts still remain in the results. In contrast, our method can remove camera motion blur without ring artifacts and restore the light streaks caused by the saturated pixels to the latent light sources. 

\begin{table*}[htbp]
\renewcommand{\arraystretch}{1.3}
\renewcommand\tabcolsep{10.0pt}
\caption{PSNR and SSIM comparisons on Kohler’s dataset}
\label{table:kohler}
\begin{center}
\begin{tabular}{c|c|c|c|c|c|c|c}
\hline
\multirow{2}{*}{Images} & \multirow{2}{*}{Measure} & \multirow{2}{*}{SRN} & \multirow{2}{*}{DeblurGAN} & \multirow{2}{*}{IRCNN (non-blind)} & \multicolumn{3}{c}{EASRN} \\
\cline{6-8}
~ & ~ & ~ & ~ & ~ & w.o. SED & w.o. OE & full \\
\hline\hline
\multirow{2}{*}{All} & PSNR & 26.75 & 24.64 & \textcolor[rgb]{1,0,0}{28.12} & 26.64 & \textcolor[rgb]{0,0,1}{27.78} & 27.45\\
\cline{2-8}
~ & SSIM & 0.8370 & 0.7880 & \textcolor[rgb]{1,0,0}{0.9208} & 0.8471 & \textcolor[rgb]{0,0,1}{0.8547} & 0.8468 \\
\hline
\multirow{2}{*}{W.o. kernel 7,8,9} & PSNR & 28.53 & 26.26 & 29.44 & 28.63 & \textcolor[rgb]{1,0,0}{30.19} & \textcolor[rgb]{0,0,1}{29.75}\\
\cline{2-8}
~ & SSIM & 0.9100 & 0.8683 & \textcolor[rgb]{1,0,0}{0.9438} & 0.9288 & \textcolor[rgb]{0,0,1}{0.9377} & 0.9346 \\
\hline
\end{tabular}
\end{center}
\end{table*}

\begin{table}[htbp]
\renewcommand{\arraystretch}{1.3}
\renewcommand\tabcolsep{15.0pt}
\caption{Performance and efficiency comparison on the GoPro test dataset.}
\label{table:GoPro}
\begin{center}
\begin{tabular}{l|c|c|c}
\hline
Measure & SRN & DeblurGAN & EASRN \\
\hline\hline
PSNR & \textcolor[rgb]{1,0,0}{28.73} & 27.15 & 27.69 \\
\hline
SSIM & \textcolor[rgb]{1,0,0}{0.9517} & 0.8755 & 0.9354 \\
\hline
Flops & 1434 G & \textcolor[rgb]{1,0,0}{678 G} & 984 G \\
\hline
Time & 0.33 s & 0.62 s & \textcolor[rgb]{1,0,0}{0.22 s} \\
\hline
\end{tabular}
\end{center}
\end{table}

\subsection{Comprehensive Comparison}

We also performed a comprehensive comparison among the proposed EASRN and both rule-based and learning-based state-of-the-art deblurring approaches on the GoPro dataset~\cite{nah2017deep}, Kohler’s dataset~\cite{kohler2012recording}, and Lai’s dataset~\cite{lai2016comparative}. SRN~\cite{tao2018scale} and DeblurGAN~\cite{kupyn2018deblurgan} are two representative deblurring networks. Pan \textit{et al.} (Pan14)~\cite{pan2014deblurring} and Xu \textit{et al.} (Xu13)~\cite{xu2013unnatural} achieved the highest performances among the traditional deblurring methods according to Lai \textit{et al.}~\cite{lai2016comparative}. Hence, the above four blind deblurring methods are our main comparisons. For SRN~\cite{tao2018scale} and DeblurGAN~\cite{kupyn2018deblurgan}, we employed the pretrained model downloaded from the authors’ webpages. For Pan14~\cite{pan2014deblurring} and Xu13~\cite{xu2013unnatural}, we employed the deblurring results reported by Lai \textit{et al.}~\cite{lai2016comparative}.

\textbf{GoPro Dataset}
The GoPro dataset~\cite{nah2017deep} represents dynamic scene blurring, which is not exactly the same as camera motion blurring. The intent of EASRN is that blur caused by limited depth of field and dynamic scenes should remain unchanged after deblurring. We randomly generated 220 synthetically blurred images from 11 GoPro videos as our test images. Table~\ref{table:GoPro} summarizes the performance and efficiency of SRN~\cite{tao2018scale}, DeblurGAN~\cite{kupyn2018deblurgan} and the proposed method. The deblurring performance of EASRN lies between that of SRN~\cite{tao2018scale} and DeblurGAN~\cite{kupyn2018deblurgan} according to the ranked PSNR and SSIM results. Considering that SRN~\cite{tao2018scale} and DeblurGAN~\cite{kupyn2018deblurgan} were trained on the GoPro dataset~\cite{nah2017deep} but our method was not, these deblurring results are acceptable. Moreover, the computational time cost of our method on a $1080\times 720$ image is only 0.22 s, which is the fastest among these methods.

\textbf{Kohler's dataset}
Kohler’s dataset~\cite{kohler2012recording} contains images that represent camera motion blurring. We employed the evaluation code from Kohler’s webpage to calculate the PSNR and SSIM. In addition to SRN~\cite{tao2018scale} and DeblurGAN~\cite{kupyn2018deblurgan}, we also tested the nonblind deblurring IRCNN~\cite{zhang2017learning} and two variants of our methods: EASRN without SED loss and EASRN trained on the dataset without OE to form a more detailed comparison. Notably, all the blind deblurring results of cases with blur kernels 7, 8, and 9 on Kohler’s dataset~\cite{kohler2012recording} are seriously degraded. We also calculated the average PSNR and SSIM excluding cases with blur kernels 7, 8, and 9 because the PSNR and SSIM values are distorted when the evaluated images are far from the ground truth. Table~\ref{table:kohler} summarizes the PSNR and SSIM results on Kohler’s dataset~\cite{kohler2012recording}. The deblurring performance of EASRN is better than that of SRN~\cite{tao2018scale} and DeblurGAN~\cite{kupyn2018deblurgan} on the dataset. Our training dataset uses a generation approach similar to that of Kohler’s dataset~\cite{kohler2012recording}; the other two models are trained on the GoPro dataset~\cite{nah2017deep}. On the dataset without kernels 7, 8, and 9, EASRN's performance is close to that of the nonblind deblurring IRCNN model~\cite{zhang2017learning}. Because Kohler’s dataset~\cite{kohler2012recording} does not include saturated blurred images, the EASRN trained without the OE dataset performs better on this dataset than does the one trained with the OE dataset.

\begin{figure*}[t]
\begin{center}
   \includegraphics[width=1\linewidth]{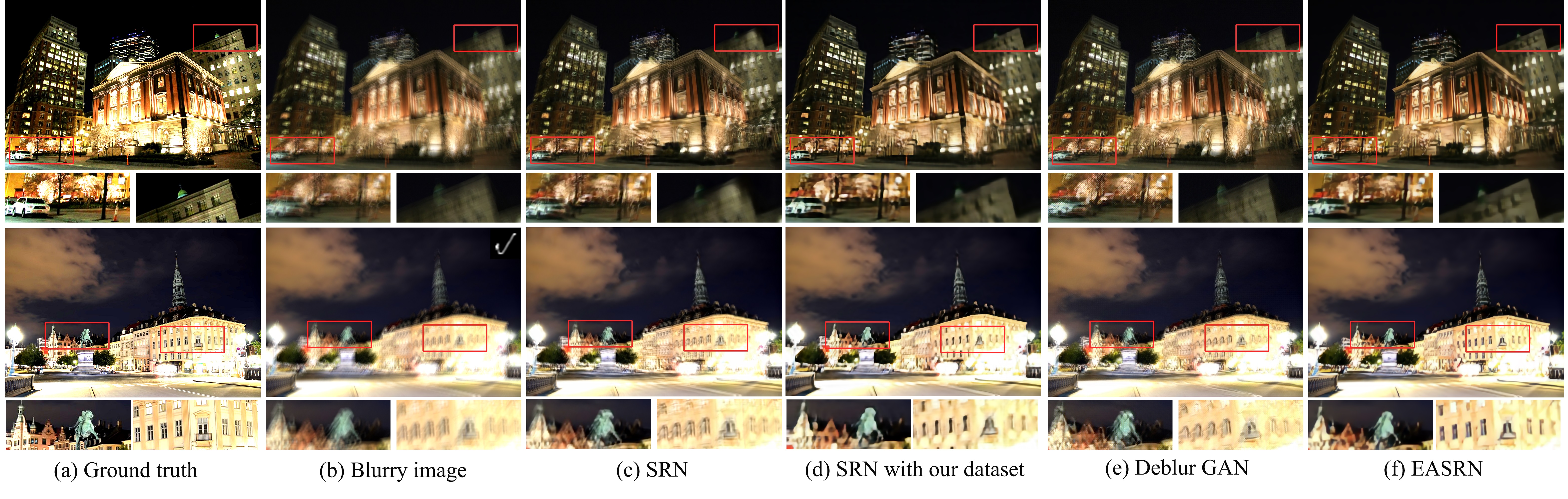}
\end{center}
   \caption{The saturated deblurring cases on the synthetic part of Lai’s dataset.}
\label{fig:lai_syn}
\end{figure*}

\begin{table}[!t]
\renewcommand{\arraystretch}{1.3}
\renewcommand\tabcolsep{9.0pt}
\caption{Performance comparisons on the synthetic portion of Lai’s dataset.}
\label{table:lai_syn}
\begin{center}
\begin{tabular}{l|c|c|c|c}
\hline
\multirow{2}{*}{Method} & \multicolumn{2}{|c|}{uniform} & \multicolumn{2}{c}{nonuniform}\\
\cline{2-5}
~ & VIF & IFC & VIF & IFC\\
\hline\hline
Pan14 & 0.0842 & 0.6260 & 0.3458 & 2.5975 \\
\hline
Xu13 & 0.0586 & 0.3659 & 0.2527 & 1.7812 \\
\hline
DeblurGAN & \textcolor[rgb]{0,0,1}{0.1245} & \textcolor[rgb]{0,0,1}{0.9430} & 0.3539 & 2.6563 \\
\hline
SRN & 0.1072 & 0.8090 & 0.3373 & 2.4884 \\
\hline
SRN with our dataset & 0.1196 & 0.9005 & \textcolor[rgb]{0,0,1}{0.3604} & \textcolor[rgb]{0,0,1}{2.7219} \\
\hline
EASRN & \textcolor[rgb]{1,0,0}{0.1409} & \textcolor[rgb]{1,0,0}{0.9775} & \textcolor[rgb]{1,0,0}{0.3851} & \textcolor[rgb]{1,0,0}{2.9063} \\
\hline
\end{tabular}
\end{center}
\end{table}

\textbf{Lai's Dataset}
Lai’s dataset~\cite{lai2016comparative} emphasizes real-world blurring. In the synthetic portion, 25 sharp images are degraded by 4 kernels/gyro. series to obtain 100 uniform/nonuniform blurred images. Moreover, 20\% of the blurred images are saturated. Hence, we additionally tested SRN with our dataset to validate the effect of our dataset. 

For the tests with the synthetic portion of Lai’s dataset~\cite{lai2016comparative}, we selected VIF~\cite{sheikh2006image} and IFC~\cite{sheikh2005information} as the evaluation indexes. According to Lai \textit{et al.}~\cite{lai2016comparative}, VIF emphasizes image edges, while IFC focuses on high-quality details. VIF and IFC are more highly correlated with subjective visualization results than PSNR and SSIM. Table~\ref{table:lai_syn} summarizes the performance comparisons on the synthetic part of Lai’s dataset~\cite{lai2016comparative}. EASRN obviously performs better than the other deblurring methods listed in Table~\ref{table:lai_syn}. The proposed dataset is beneficial for saturated blurred image deblurring. Fig.~\ref{fig:lai_syn} shows the saturated deblurring cases on the synthetic part of Lai’s dataset~\cite{lai2016comparative}. The upper image is a nonuniform blurred image, and the bottom image is a uniform blurred image. Fig.~\ref{fig:lai_syn}(a) shows the ground truth images; Fig.~\ref{fig:lai_syn}(b) shows the blurred images Fig.~\ref{fig:lai_syn}(c) shows the SRN~\cite{tao2018scale} results; Fig.~\ref{fig:lai_syn}(d) shows the SRN~\cite{tao2018scale} results on our dataset; Fig.~\ref{fig:lai_syn}(e) shows the DeblurGAN [18] results; and Fig.~\ref{fig:lai_syn}(f) shows the EASRN results. Comparing Fig.~\ref{fig:lai_syn}(c) and Fig.~\ref{fig:lai_syn}(d), SRN with our dataset gains the capability of restoring the saturated images, while the SRN trained with the GoPro dataset~\cite{nah2017deep} cannot distinguish the real blur kernel under the interference of outliers. Comparing Fig.~\ref{fig:lai_syn}(f) with the other deblurring results, the EASRN results have a higher detail contrast and fewer ringing artifacts. In the nonuniform blurred case, we select the lower-left and upper-right patches of the image and specifically focus on them. These two patches have different blur kernels. Both patches are well deblurred in EASRN, which indicates its capability for nonuniform deblurring.

\begin{table}
\renewcommand{\arraystretch}{1.5}
\renewcommand\tabcolsep{10.0pt}
\caption{Half-subjective image quality comparison  on the real-world part of Lai's dataset.}
\label{table:lai_real}
\begin{center}
\begin{tabular}{l|c}
\hline
Method & Average quality  \\
\hline\hline
DeblurGAN &  -10.71\\
\hline
Pan14 & -10.78\\
\hline
Xu13 & \textcolor[rgb]{0,0,1}{-10.58}\\
\hline
SRN with our dataset & -11.44\\
\hline
SRN & -11.79 \\
\hline
EASRN & \textcolor[rgb]{1,0,0}{-9.97} \\
\hline
\end{tabular}
\end{center}
\end{table}

\begin{figure}[t]
\begin{center}
   \includegraphics[width=1\linewidth]{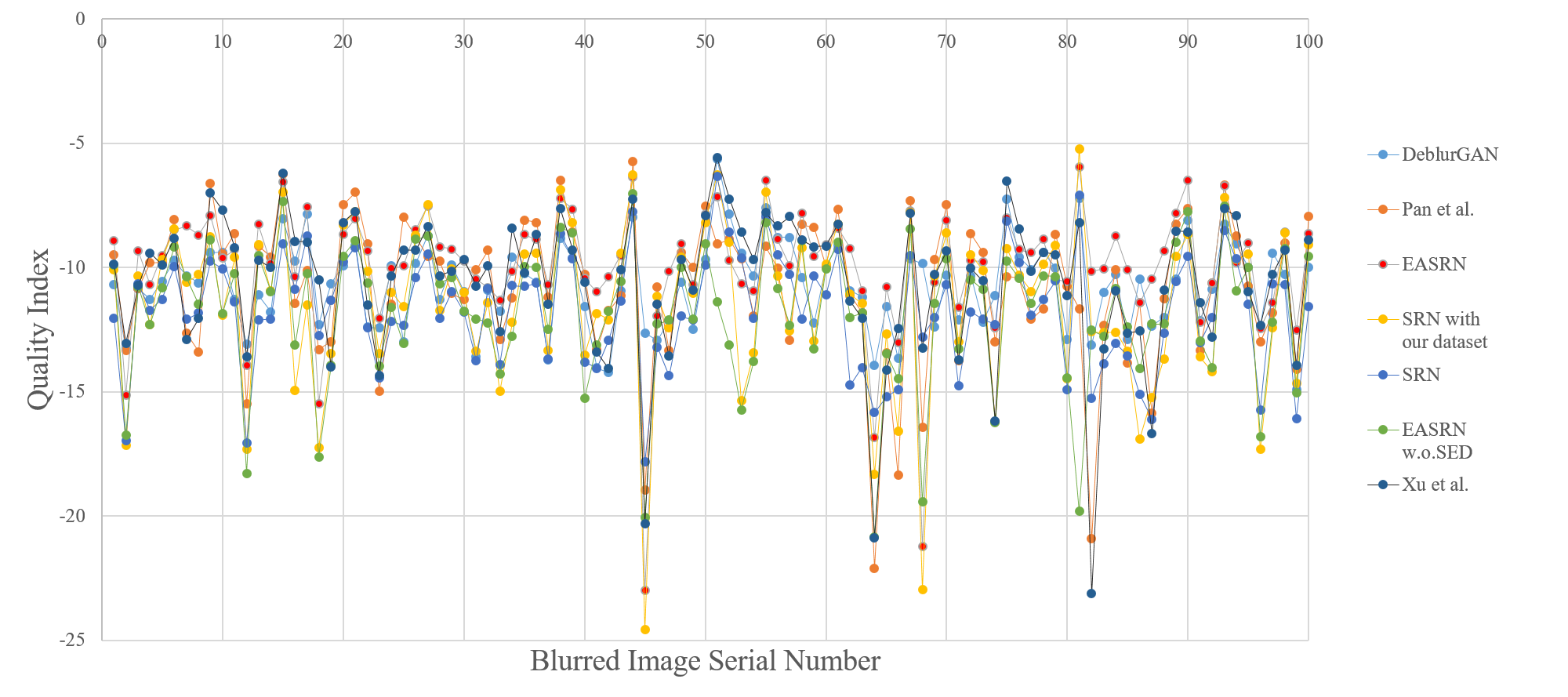}
\end{center}
   \caption{The image quality index results on the real-world part of Lai’s dataset.}
\label{fig:index}
\end{figure}

\begin{figure*}[t]
\begin{center}
   \includegraphics[width=1\linewidth]{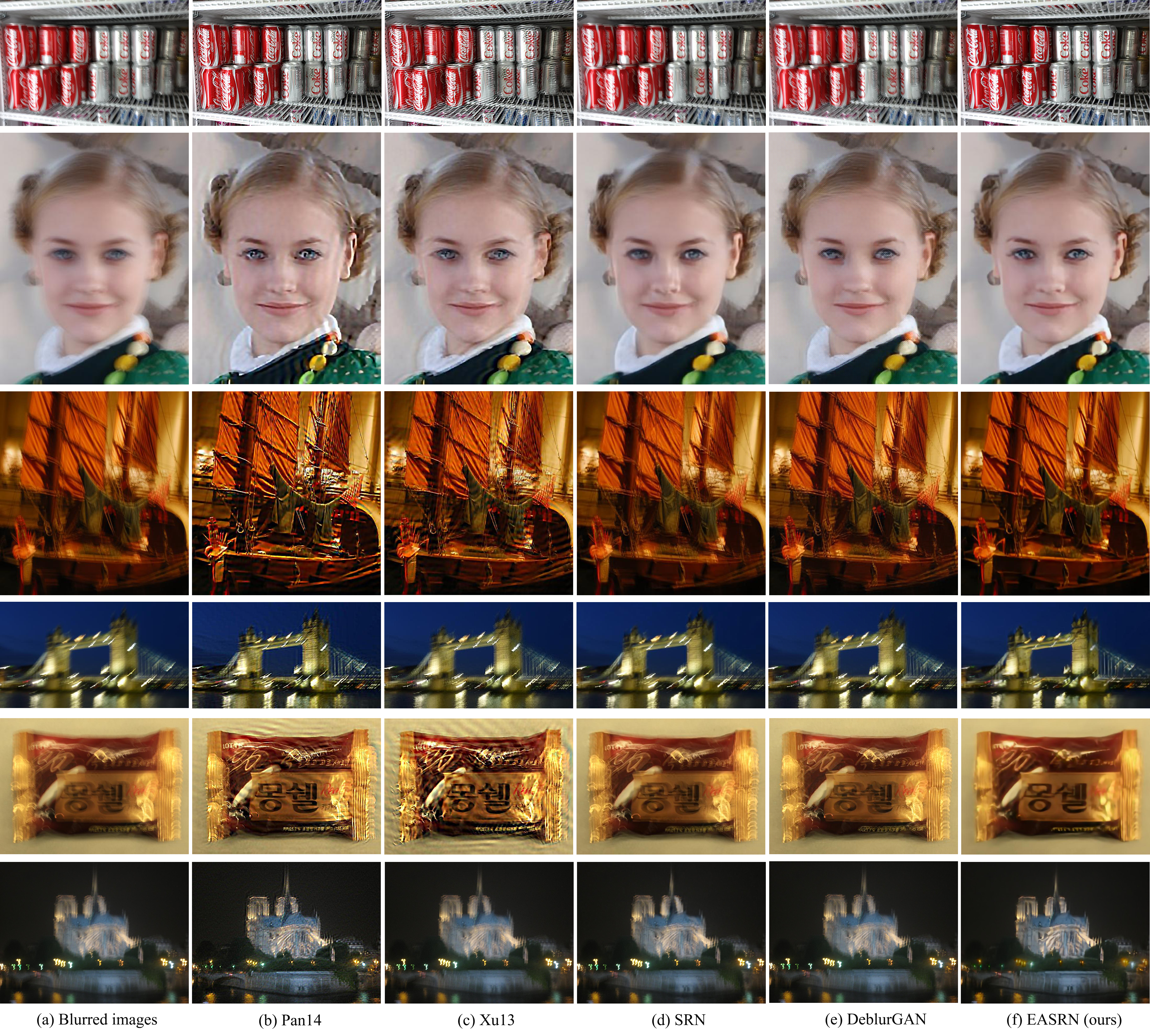}
\end{center}
   \caption{The typical deblurring cases on the real-world part of Lai’s dataset.}
\label{fig:lai_real}
\end{figure*}

The real-world part of Lai’s dataset~\cite{lai2016comparative} consists of 100 real-world blurred images, of which 27 are saturated images. These images cover various types of scenes and have different qualities and resolutions. We select a half-subjective no-reference evaluation method proposed by Liu \textit{et al.}~\cite{liu2013no} to validate the performance of EASRN. Using this method, a higher image quality index refers to a better deblurring effect. Table~\ref{table:lai_real} summarizes the half-subjective image quality comparison on the real-world part of Lai’s dataset~\cite{lai2016comparative}. EASRN performs the best among the tested methods. Fig.~\ref{fig:index} shows the specific scores on 100 images. On the entire dataset, the quality indexes of EASRN (red spots in Fig.~\ref{fig:index}) occupy 40\% of the top 1 scores and 84\% of the top 3, which demonstrates the robustness of EASRN. Fig.~\ref{fig:lai_real} shows some typical deblurring cases on the real-world part of Lai’s dataset~\cite{lai2016comparative}. From top to bottom, the images are named “Coke”, “Face”, “Boat”, “Bridge”, “Candy” and “Notredame”. The images Face and Boat are normally exposed, while the others are partially overexposed. Due to the closeup imaging and the variable depths of field, the images Face and Boat have nonuniform blur. The rule-based deblurring approaches, such as Pan14~\cite{pan2014deblurring} and Xu13~\cite{xu2013unnatural} are not good at handling nonuniform blur; these methods result in ringing artifacts. The CNN deblurring methods based on the GoPro dataset, such as SRN~\cite{tao2018scale} and DeblurGAN~\cite{kupyn2018deblurgan}, emphasize partial deblurring, which results in the failure of large blur restoration in complex texture regions. For the saturated cases, only Pan14~\cite{pan2014deblurring} has the capability to deblur images with outliers; however, its limited deconvolution performance causes the deblurring results to have obvious ringing artifacts. Compared with the other methods, the deblurring results of EASRN have sharper salient edges and include more high-frequency details, fewer ringing artifacts, and less noise.

\section{Conclusion}

In this paper, we proposed an edge-aware scale-recurrent neural network, called EASRN, to perform camera motion deblurring with outliers. Benefiting from separate subnets and a scale-recurrent approach, EASRN can handle more complicated blurring situations. We successfully solved the saturation problem in deblurring by devising a novel dataset generation method, and proposed a salient edge detection neural network to assist in constructing a loss function to suppress ring artifacts. When equipped with the three contributions mentioned above, EASRN obtains better image quality in the deblurred results than do other state-of-the-art methods.

\section*{Acknowledgment}
This work was supported by the National Science Foundation of China (No. 61975175) and the Basic Research Projects (No. 2018110C081).

\ifCLASSOPTIONcaptionsoff
  \newpage
\fi



\bibliographystyle{IEEEtran}
\bibliography{reference}

\begin{thebibliography}{10}
\providecommand{\url}[1]{#1}
\csname url@samestyle\endcsname
\providecommand{\newblock}{\relax}
\providecommand{\bibinfo}[2]{#2}
\providecommand{\BIBentrySTDinterwordspacing}{\spaceskip=0pt\relax}
\providecommand{\BIBentryALTinterwordstretchfactor}{4}
\providecommand{\BIBentryALTinterwordspacing}{\spaceskip=\fontdimen2\font plus
\BIBentryALTinterwordstretchfactor\fontdimen3\font minus
  \fontdimen4\font\relax}
\providecommand{\BIBforeignlanguage}[2]{{%
\expandafter\ifx\csname l@#1\endcsname\relax
\typeout{** WARNING: IEEEtran.bst: No hyphenation pattern has been}%
\typeout{** loaded for the language `#1'. Using the pattern for}%
\typeout{** the default language instead.}%
\else
\language=\csname l@#1\endcsname
\fi
#2}}
\providecommand{\BIBdecl}{\relax}
\BIBdecl

\bibitem{shan2008high}
Q.~Shan, J.~Jia, and A.~Agarwala, ``High-quality motion deblurring from a
  single image,'' \emph{Acm transactions on graphics (tog)}, vol.~27, no.~3,
  p.~73, 2008.

\bibitem{cho2011handling}
S.~Cho, J.~Wang, and S.~Lee, ``Handling outliers in non-blind image
  deconvolution,'' in \emph{2011 International Conference on Computer
  Vision}.\hskip 1em plus 0.5em minus 0.4em\relax IEEE, 2011, pp. 495--502.

\bibitem{whyte2014deblurring}
O.~Whyte, J.~Sivic, and A.~Zisserman, ``Deblurring shaken and partially
  saturated images,'' \emph{International journal of computer vision}, vol.
  110, no.~2, pp. 185--201, 2014.

\bibitem{bar2006image}
L.~Bar, N.~Kiryati, and N.~Sochen, ``Image deblurring in the presence of
  impulsive noise,'' \emph{International Journal of Computer Vision}, vol.~70,
  no.~3, pp. 279--298, 2006.

\bibitem{fergus2006removing}
R.~Fergus, B.~Singh, A.~Hertzmann, S.~T. Roweis, and W.~T. Freeman, ``Removing
  camera shake from a single photograph,'' in \emph{ACM transactions on
  graphics (TOG)}, vol.~25, no.~3.\hskip 1em plus 0.5em minus 0.4em\relax ACM,
  2006, pp. 787--794.

\bibitem{krishnan2009fast}
D.~Krishnan and R.~Fergus, ``Fast image deconvolution using hyper-laplacian
  priors,'' in \emph{Advances in neural information processing systems}, 2009,
  pp. 1033--1041.

\bibitem{pan2014deblurring}
J.~Pan, Z.~Hu, Z.~Su, and M.-H. Yang, ``Deblurring text images via
  l0-regularized intensity and gradient prior,'' in \emph{Proceedings of the
  IEEE Conference on Computer Vision and Pattern Recognition}, 2014, pp.
  2901--2908.

\bibitem{pan2016blind}
J.~Pan, D.~Sun, H.~Pfister, and M.-H. Yang, ``Blind image deblurring using dark
  channel prior,'' in \emph{Proceedings of the IEEE Conference on Computer
  Vision and Pattern Recognition}, 2016, pp. 1628--1636.

\bibitem{xu2010two}
L.~Xu and J.~Jia, ``Two-phase kernel estimation for robust motion deblurring,''
  in \emph{European conference on computer vision}.\hskip 1em plus 0.5em minus
  0.4em\relax Springer, 2010, pp. 157--170.

\bibitem{pan2016robust}
J.~Pan, Z.~Lin, Z.~Su, and M.-H. Yang, ``Robust kernel estimation with outliers
  handling for image deblurring,'' in \emph{Proceedings of the IEEE Conference
  on Computer Vision and Pattern Recognition}, 2016, pp. 2800--2808.

\bibitem{hu2016image}
Z.~Hu, L.~Yuan, S.~Lin, and M.-H. Yang, ``Image deblurring using smartphone
  inertial sensors,'' in \emph{Proceedings of the IEEE Conference on Computer
  Vision and Pattern Recognition}, 2016, pp. 1855--1864.

\bibitem{joshi2008psf}
N.~Joshi, R.~Szeliski, and D.~J. Kriegman, ``Psf estimation using sharp edge
  prediction,'' in \emph{2008 IEEE Conference on Computer Vision and Pattern
  Recognition}.\hskip 1em plus 0.5em minus 0.4em\relax IEEE, 2008, pp. 1--8.

\bibitem{sun2015learning}
J.~Sun, W.~Cao, Z.~Xu, and J.~Ponce, ``Learning a convolutional neural network
  for non-uniform motion blur removal,'' in \emph{Proceedings of the IEEE
  Conference on Computer Vision and Pattern Recognition}, 2015, pp. 769--777.

\bibitem{xiao2016learning}
L.~Xiao, J.~Wang, W.~Heidrich, and M.~Hirsch, ``Learning high-order filters for
  efficient blind deconvolution of document photographs,'' in \emph{European
  Conference on Computer Vision}.\hskip 1em plus 0.5em minus 0.4em\relax
  Springer, 2016, pp. 734--749.

\bibitem{zhang2017learning}
K.~Zhang, W.~Zuo, S.~Gu, and L.~Zhang, ``Learning deep cnn denoiser prior for
  image restoration,'' in \emph{Proceedings of the IEEE conference on computer
  vision and pattern recognition}, 2017, pp. 3929--3938.

\bibitem{tao2018scale}
X.~Tao, H.~Gao, X.~Shen, J.~Wang, and J.~Jia, ``Scale-recurrent network for
  deep image deblurring,'' in \emph{Proceedings of the IEEE Conference on
  Computer Vision and Pattern Recognition}, 2018, pp. 8174--8182.

\bibitem{nah2017deep}
S.~Nah, T.~Hyun~Kim, and K.~Mu~Lee, ``Deep multi-scale convolutional neural
  network for dynamic scene deblurring,'' in \emph{Proceedings of the IEEE
  Conference on Computer Vision and Pattern Recognition}, 2017, pp. 3883--3891.

\bibitem{kupyn2018deblurgan}
O.~Kupyn, V.~Budzan, M.~Mykhailych, D.~Mishkin, and J.~Matas, ``Deblurgan:
  Blind motion deblurring using conditional adversarial networks,'' in
  \emph{Proceedings of the IEEE Conference on Computer Vision and Pattern
  Recognition}, 2018, pp. 8183--8192.

\bibitem{liu2013no}
Y.~Liu, J.~Wang, S.~Cho, A.~Finkelstein, and S.~Rusinkiewicz, ``A no-reference
  metric for evaluating the quality of motion deblurring.'' \emph{ACM Trans.
  Graph.}, vol.~32, no.~6, pp. 175--1, 2013.

\bibitem{hirsch2011fast}
M.~Hirsch, C.~J. Schuler, S.~Harmeling, and B.~Sch{\"o}lkopf, ``Fast removal of
  non-uniform camera shake,'' in \emph{2011 International Conference on
  Computer Vision}.\hskip 1em plus 0.5em minus 0.4em\relax IEEE, 2011, pp.
  463--470.

\bibitem{boracchi2012modeling}
G.~Boracchi and A.~Foi, ``Modeling the performance of image restoration from
  motion blur,'' \emph{IEEE Transactions on Image Processing}, vol.~21, no.~8,
  pp. 3502--3517, 2012.

\bibitem{rudin1992nonlinear}
L.~I. Rudin, S.~Osher, and E.~Fatemi, ``Nonlinear total variation based noise
  removal algorithms,'' \emph{Physica D: nonlinear phenomena}, vol.~60, no.
  1-4, pp. 259--268, 1992.

\bibitem{richardson1972bayesian}
W.~H. Richardson, ``Bayesian-based iterative method of image restoration,''
  \emph{JoSA}, vol.~62, no.~1, pp. 55--59, 1972.

\bibitem{gong2017motion}
D.~Gong, J.~Yang, L.~Liu, Y.~Zhang, I.~Reid, C.~Shen, A.~Van Den~Hengel, and
  Q.~Shi, ``From motion blur to motion flow: a deep learning solution for
  removing heterogeneous motion blur,'' in \emph{Proceedings of the IEEE
  Conference on Computer Vision and Pattern Recognition}, 2017, pp. 2319--2328.

\bibitem{gao2019dynamic}
H.~Gao, X.~Tao, X.~Shen, and J.~Jia, ``Dynamic scene deblurring with parameter
  selective sharing and nested skip connections,'' in \emph{Proceedings of the
  IEEE Conference on Computer Vision and Pattern Recognition}, 2019, pp.
  3848--3856.

\bibitem{dong2017blind}
J.~Dong, J.~Pan, Z.~Su, and M.-H. Yang, ``Blind image deblurring with outlier
  handling,'' in \emph{Proceedings of the IEEE International Conference on
  Computer Vision}, 2017, pp. 2478--2486.

\bibitem{hu2014deblurring}
Z.~Hu, S.~Cho, J.~Wang, and M.-H. Yang, ``Deblurring low-light images with
  light streaks,'' in \emph{Proceedings of the IEEE Conference on Computer
  Vision and Pattern Recognition}, 2014, pp. 3382--3389.

\bibitem{he2016deep}
K.~He, X.~Zhang, S.~Ren, and J.~Sun, ``Deep residual learning for image
  recognition,'' in \emph{Proceedings of the IEEE conference on computer vision
  and pattern recognition}, 2016, pp. 770--778.

\bibitem{SzegedyGoing}
C.~Szegedy, W.~Liu, Y.~Jia, P.~Sermanet, S.~Reed, D.~Anguelov, D.~Erhan,
  V.~Vanhoucke, and A.~Rabinovich, ``Going deeper with convolutions.''

\bibitem{maas2013rectifier}
A.~L. Maas, A.~Y. Hannun, and A.~Y. Ng, ``Rectifier nonlinearities improve
  neural network acoustic models,'' in \emph{Proc. icml}, vol.~30, no.~1, 2013,
  p.~3.

\bibitem{JohnsonPerceptual}
J.~Johnson, A.~Alahi, and L.~Fei-Fei, ``Perceptual losses for real-time style
  transfer and super-resolution.''

\bibitem{xie2015holistically}
S.~Xie and Z.~Tu, ``Holistically-nested edge detection,'' in \emph{Proceedings
  of the IEEE international conference on computer vision}, 2015, pp.
  1395--1403.

\bibitem{kohler2012recording}
R.~K{\"o}hler, M.~Hirsch, B.~Mohler, B.~Sch{\"o}lkopf, and S.~Harmeling,
  ``Recording and playback of camera shake: Benchmarking blind deconvolution
  with a real-world database,'' in \emph{European conference on computer
  vision}.\hskip 1em plus 0.5em minus 0.4em\relax Springer, 2012, pp. 27--40.

\bibitem{lai2016comparative}
W.-S. Lai, J.-B. Huang, Z.~Hu, N.~Ahuja, and M.-H. Yang, ``A comparative study
  for single image blind deblurring,'' in \emph{Proceedings of the IEEE
  Conference on Computer Vision and Pattern Recognition}, 2016, pp. 1701--1709.

\bibitem{xu2013unnatural}
L.~Xu, S.~Zheng, and J.~Jia, ``Unnatural l0 sparse representation for natural
  image deblurring,'' in \emph{Proceedings of the IEEE conference on computer
  vision and pattern recognition}, 2013, pp. 1107--1114.

\bibitem{sheikh2006image}
H.~R. Sheikh and A.~C. Bovik, ``Image information and visual quality,''
  \emph{IEEE Transactions on image processing}, vol.~15, no.~2, pp. 430--444,
  2006.

\bibitem{sheikh2005information}
H.~R. Sheikh, A.~C. Bovik, and G.~De~Veciana, ``An information fidelity
  criterion for image quality assessment using natural scene statistics,''
  \emph{IEEE Transactions on image processing}, vol.~14, no.~12, pp.
  2117--2128, 2005.

\end{thebibliography}






\end{document}